\documentclass[runningheads]{llncs}
\usepackage{amsmath,amssymb,amsfonts}
\usepackage{algorithmic}
\usepackage{graphicx}
\usepackage{tabularx}
\usepackage{textcomp}
\usepackage{xcolor}
\usepackage{subfigure}
\usepackage{graphicx}
\usepackage[ruled,vlined]{algorithm2e}
\usepackage{multirow}
\usepackage{framed}
\usepackage{enumitem}
\usepackage{sourcecodepro}
\usepackage[T1]{fontenc}
\usepackage{soul}
\usepackage{tikz}
\usepackage{float}
\usepackage{array}
\usepackage{url}
\usepackage{ulem}
\usepackage{booktabs}

\usetikzlibrary{positioning}
\tikzset{>=stealth}

\definecolor{LemonChiffon}{RGB}{255,250,205}

\newcommand{\tikzmark}[3][]
{\tikz[remember picture, baseline]
 \node [anchor=base,#1](#2) {#3};}

\definecolor{grey}{RGB}{220, 220, 220}
\definecolor{darkgrey}{RGB}{170, 170, 170}

\makeatletter
\newcommand{\thickhline}{%
    \noalign {\ifnum 0=`}\fi \hrule height 1pt
    \futurelet \reserved@a \@xhline
}

\begin{document}

\title{Decentralized Threshold Signatures with Dynamically Private Accountability}

\author{Meng Li\inst{1}\and
Mingwei Zhang\inst{1}\and
Qing Wang\inst{1}\and
Hanni Ding\inst{1}\and\\
Weizhi Meng\inst{2}\and
Liehuang Zhu\inst{3}\thanks{Liehuang Zhu and Zijian Zhang are corresponding authors.}\and
Zijian Zhang\inst{3}$^{\star}$\and
Xiaodong Lin\inst{4}
}

\authorrunning{}

\institute{Hefei University of Technology, China
\and
Technical University of Denmark, Denmark
\and
Beijing Institute of Technology, China
\and
University of Guelph, Canada
\\
\email{mengli@hfut.edu.cn, \{mwzhang, qingwang, hanniding\}@mail.hfut.edu.cn,\\weme@dtu.dk, \{liehuangz, zhangzijian\}@bit.edu.cn, xlin08@uoguelph.ca}
}

\maketitle

\begin{abstract}
Threshold signatures are a fundamental cryptographic primitive used in many practical applications.
As proposed by Boneh and Komlo (CRYPTO'22), TAPS is a threshold signature that is a hybrid of privacy and accountability. It enables a combiner to combine $t$ signature shares while revealing nothing about the threshold $t$ or signing quorum to the public and asks a tracer to track a signature to the quorum that generates it.
However, TAPS has three disadvantages: it 1) structures upon a centralized model, 2) assumes that both combiner and tracer are honest, and 3) leaves the tracing unnotarized and static.

In this work, we introduce Decentralized, Threshold, dynamically Accountable and Private Signature (DeTAPS) that provides decentralized combining and tracing, enhanced privacy against untrusted combiners (tracers), and notarized and dynamic tracing.
Specifically, we adopt Dynamic Threshold Public-Key Encryption (DTPKE) to dynamically notarize the tracing process, design non-interactive zero-knowledge proofs to achieve public verifiability of notaries, and utilize the Key-Aggregate Searchable Encryption to bridge TAPS and DTPKE so as to awaken the notaries securely and efficiently.
In addition, we formalize the definitions and security requirements for DeTAPS. Then we present a generic construction and formally prove its security and privacy.
To evaluate the performance, we build a prototype based on SGX2 and Ethereum.
\end{abstract}

\section{Introduction}
\subsection{Background}
Threshold signatures~\cite{TS89,TS00} allow a group of $n$ parties to sign a message if no less than $t$ parties participate in the signing process. They are a crucial tool for many practical applications~\cite{EUROCRYPT01,ASIACRYPT21,CCS221}.
Among them, there are two types of threshold signatures standing out: Accountable Threshold Signature (ATS) and Private Threshold Signature (PTS).
An ATS $\sigma$ on a message $m$ reveals the identity of all $t$ signers who co-generated the signature~\cite{ATS1,ATS2}.
A PTS $\sigma$ on a message $m$ reveals nothing about $t$ or the quorum of $t$ original signers~\cite{PTS1,PTS2}.
Besides unforgeability, these two signatures offer complete accountability and complete privacy for the signing quorum, respectively.

\subsection{Existing Work}
A recent work Threshold, Accountable, and Private Signature (TAPS)~\cite{TAPS22} proposed by Boneh and Komlo (CRYPTO'22) has achieved both accountability and privacy. In TAPS, each signer from a quorum of $t$ signers $\mathcal{S}$, holding a private key $sk$, generates a signature share $\sigma_i$; a combiner holding a combining key $sk_c$ uses $\{\sigma_i\}_{i=1}^n$ to generate a complete signature $\sigma$; a tracer with a tracing key $sk_t$ can trace a signature to the quorum that generates it.
The benefits of TAPS are remarkable: the signing group keeps the $sk_t$ secret so that $t$ and $\mathcal{S}$ remain private from the public, but the $t$ signers are accountable in case of misbehaviors~\cite{ESORICS1}.

\subsection{Motivations and New Goals}
The observations on TAPS lead to our \textbf{four motivations}.
\textbf{M1. Centralized combining and tracing}. The role of combiner and tracer is important to generating and tracing a complete signature. However, the centralized setting is prone to a single point of failure.
\textbf{M2. Untrusted combiner}. The combining key $sk_c$ is kept by the only combiner that could be untrusted, e.g., lose or leak the key. The threshold $t$ is also exposed to the combiner. As designed in TAPS, $t$ is part of privacy and is hidden from the public. Therefore, we take the privacy one step further by assuming an untrusted combiner.
\textbf{M3. Untrusted tracer}. Similarly, the tracing key $sk_t$ is kept by an untrusted tracer and $t$ is exposed.
\textbf{M4. Unnotarized and static tracing}. The tracing key $sk_t$ is kept by the sole tracer that can use $sk_t$ to recover any quorum of $t$ signers.
We argue that the tracing process is a sensitive process that should be notarized by a dynamic and relevant group of notaries, i.e., $t'$ notaries or witnesses~\cite{SP16}.
For example, a company's assets are frozen due to engaging in criminal activities and such a sanction should be approved by $t'$ authorities such as police department and finance department.
Meanwhile, the value of $t'$ varies according to the matter and relevant authorities.
The idea resembles the one in threshold encryption where a ciphertext can only  be decrypted when at least $t'$ users cooperate~\cite{DTPKE08}.

These four motivations have driven us to provide enhanced security and privacy in threshold signatures using a decentralized approach, i.e., decentralized threshold signatures with dynamically private accountability. Namely, we have four new goals as follows.
\textbf{G1. Enhanced security against a single point of failure}. The threshold signature system should be secure in a decentralized manner such that one (a small number of) combiner/tracer's breakdown does not affect the whole system.
\textbf{G2. Enhanced privacy against untrusted combiners}. The threshold signature system is privacy-preserving during the signing process. Specifically, not only the quorum of $t$ original singers, but $t$, $sk_t$, and $t'$ are hidden from combiners.
\textbf{G3. Enhanced privacy against untrusted tracers}. The threshold signature system is privacy-preserving during the tracing process. To be specific, $t$, $sk_c$, and $t'$ are hidden from tracers.
\textbf{G4. Notarized and dynamic tracing}. The tracing process should be notarized by $t'$ parties among a group of authorities. The value of $t'$ is a variable parameter, which is related to the specific tracing requirement.

\textit{Remark 1 (Privacy of $t$ signers after tracing).}
We notice that once a tracer has traced a complete signature to its $t$ signers, the signers' identities as well as $t$ are revealed to the tracer.
This looks contradictory to G3 where we protect $t$ and make G3 only applicable to the realm before tracing.
However, we can choose to protect $t$ signers from tracers (will be explained Section 4).

\vspace{-0.5cm}
\subsection{Our Approach}
To achieve the four abovementioned goals, we propose an approach as follows.
(1) We transit the centralized model of TPAS into a decentralized one by using a Consortium Blockchain (CB)~\cite{CB1,CB2} to distribute the combining and tracing capabilities. Each blockchain node can be either a combiner or a tracer such that the combiner (tracer) actually performing the combining (tracing) is determined by the underlying consensus mechanism. In this way, an adversary cannot predict such a performer to attack.
(2) We protect $t$ and $t'$ from the untrusted combiners and untrusted tracers during the combining and tracing by deploying an Trusted Execution Environment (TEE)~\cite{TEE1,SGX21,SGX22} on combiners and tracers. The combining and tracing will be conducted within a secure enclave, over which the combiners and tracers have no control over the data inside.
(3) We propose ``dynamically private accountability'', i.e., limit the tracing capability of untrusted tracers by asking another quorum of $t'$ parties to notarize the tracing process. We denote this quorum as $\mathcal{N} = \{N_1, N_2, \cdots, N_{t'}\}$. Specifically, the tracer can only trace from a complete signature to its $t$ signers only if there are $t'$ notaries allow it. This is realized by adopting Dynamic Threshold Public-Key Encryption (DTPKE)~\cite{DTPKE08} to designate $t'$ notaries for the tracing process.

In summary, we introduce a new type of threshold signature scheme, called DeTAPS, that provides dynamic accountability while maintaining full privacy for the signing quorum and notarizing quorum.
A \textbf{Decentralized, Threshold, dynamically Accountable and Private Signature} scheme, or simply \textbf{DeTAPS}, works as follows:
(i) a key generation procedure generates a public key $pk$ and $n$ private keys $\{sk_1, sk_2, \cdots,sk_n\}$, a combining key $sk_c$, and a tracing key $sk_t$,
(ii) a signing protocol among a quorum of $t$ signers and a combiner generate a signature $\sigma$ on a message $m$,
(iii) a signature verification algorithm takes as input $pk$, $m$, and $\sigma$ and outputs true or false, and
(iv) a tracing algorithm takes as input $sk_t$, $m$, and $\sigma$, and outputs the original quorum of $t$ signers.
For security model, we assume that the combiners and tracers are malicious, which are not allowed to know $t$ or $t'$.
We define the precise syntax for the DeTAPS scheme and the security requirements in Section 3.

\subsection{Technical Challenges}
Given the general approach, we are still faced with three technical challenges when constructing DeTAPS:

\textbf{C1}. How to securely select the $t'$ notaries while guaranteeing public verifiability?
In this work, we ask the $t$ signers to choose $t'$ notaries whose identities are kept secret.
In the meantime, we have to guarantee public verifiability of the $t'$ notaries, i.e., there are enough authenticated notaries selected by the $t$ signers during combining.

\textbf{C2}. How to securely awaken the $t'$ notaries to the call for partially decryption of encrypted threshold signatures when necessary?
There are several technical candidates for solving this problem.
(1) Encrypt-and-Decrypt: It is workable, but is time consuming and clumsy.
(2) Private Set Intersection (PSI)~\cite{PSI1,PSI2}: It provides strong security but requires more than one interaction, which results in high costs.
(3) Attribute-Based Encryption (ABE)~\cite{CCS06,ESORICS2}: It achieves fine-grained access control but incurs high computational costs.

\textbf{C3}. How to allow a notary to efficiently locate the encrypted signatures related to himself from all the ciphertexts on the CB?
Some technical candidates are as follows.
(1) Indistinguishable Bloom Filter (IBF)~\cite{IBF1,IBF2}: It is efficient but needs to share a set of keys between signers and notaries.
(2) Designated Verifier Signature~\cite{DVS03,DVS07}: It requires additional signing by the signers and it cannot provide confidentiality.

To tackle C1, we design Non-Interactive Zero Knowledge Proofs (NIZKPs) to enable the public to verify the $t'$ notaries in a secure manner.
To overcome C2 and C3, we utilize the Key-Aggregate Searchable Encryption (KASE)~\cite{KASE16} as a bridge between TAPS and DTPKE to reconcile   security and efficiency.

We provide some details on how we construct DeTAPS.
\textbf{Setup}. We assume that any quorum of $t$ signers have communicated with each other via face to face or a secure channel to determine $t'$ notaries $\mathcal{N} = \{N_1, N_2, \cdots, N_{t'}\}$.
Each quorum of $t$ signers has a unique and random signer group identifier $gid\in \mathcal{G}$ in each signing period. This can be done by asking a representative of each quorum to anonymously write a random number on the blockchain. $\mathcal{G}$ will be updated in future periods.
Each notary has a pseudo-identity $pid$.
A KASE aggregation key $k_a$ is generated in the beginning for each notary.
\textbf{Sign}. Each signer of a quorum of $t$ signers generates a signature share $\sigma_i$ on the same message $m$ and sends its ciphertext to the CB.
\textbf{Combine}.
During combining, the enclave $E$ within the combiner $C$ encrypts $\sigma$ to be an encrypted threshold signature $\overline{\sigma}$ by using the combining key $sk_c$.
After combining, $E$ computes an index $ind$ of $\mathcal{N}$.
\textbf{Trace}.
Upon a tracing call, each related notary computes a trapdoor $td$ by using $k_a$ and $pid$.
The index and trapdoor are sent to a smart contract~\cite{TDSC222} that searches on $ind$ with $td$ to retrieve a matched $\overline{\sigma}$ to the requesting notary. The notary sends a partial decryption of $\overline{\sigma}$ to the CB.
Only if the designated $t'$ notaries are awaken to perform partially decryption, can a tracer $T$ trace within its enclave to the original quorum of $t$ tracers by using the tracing key $sk_t$.
In addition, the encrypted threshold signature can be verified by the public.

\subsection{Our Contributions}

\begin{itemize}
    \item We design a decentralized framework for threshold signatures to distribute the combining (tracing) capabilities to multiple combiners (tracers).
    \item We design a TEE-based execution engine to secure the combining (tracing) process against untrusted combiners (tracers).
    \item We adopt DTPKE to dynamically notarize the tracing process and integrate TAPS with DTPKE by using KASE to awaken the notaries.
    \item We formally state and prove the security and privacy of DeTAPS.
\end{itemize}

\textit{Paper Organization.} The paper is organized as follows.
Section 2 briefly reviews some preliminaries.
Section 3 formalizes the system model, security, and privacy of DeTAPS.
Section 4 describes DeTAPS.
Section 5 analyzes its security and privacy.
Section 6 evaluates the performance of DeTAPS.
Section 7 concludes this paper.

\section{Preliminaries}
In this section, we briefly review some preliminaries that work as building blocks.

\subsection{ATS}
An accountable threshold signature is a tuple of five polynomial time algorithms $(\mathsf{KeyGen}, \mathsf{KeyGen}, \mathsf{Sign}, \mathsf{Combine},\mathsf{Verify}, \mathsf{Trace})$ invoked as
\begin{gather*}
(pk, (sk_1, sk_2, \cdots, sk_n)) \leftarrow \mathsf{ATS.KeyGen}(1^{\lambda}, n, t),\\
\sigma_i\leftarrow \textsf{ATS.Sign}(sk_i, m),\ \sigma_m \leftarrow \mathsf{ATS.Combine}(pk, m, \mathcal{S}, \{\sigma_i\}_{i\in \mathcal{S}}),\\
\{0,1\}\leftarrow \mathsf{ATS.Verify}(pk, m, \sigma_m),\\
\mathcal{S}\leftarrow \mathsf{ATS.Trace}(pk, m, \sigma_m).
\end{gather*}

\noindent$\mathsf{ATS.KeyGen}:$ takes as input a security parameter $\lambda$, the number of parties $n$ and threshold $t$. It outputs a public key $pk$ and signer keys $(sk_1, sk_2, \cdots, sk_n)$.

\noindent$\mathsf{ATS.Sign}:$ takes as input a key of signer $sk_i$, the signing quorum $\mathcal{S}$ and the message $m$. It outputs a signature $\sigma_i$.

\noindent$\mathsf{ATS.Combine}:$ takes as input $pk, m, \mathcal{S}$ and the signature set $\{\sigma_i\}_{i\in \mathcal{S}}$. It outputs a ATS signature $\sigma_m$.

\noindent$\mathsf{ATS.Verify}:$ takes as input $pk, m, \sigma_m$ and outputs 0 or 1.

\noindent$\mathsf{ATS.Trace}:$ takes as input $pk, m, \sigma_m$ and outputs the signing quorum $\mathcal{S}$.

An ATS is said to be secure if it is unforgeable and accountable, i.e., if for every Probabilistic Polynomial Time (PPT) adversary $\mathcal{A}$, the function $\textbf{Adv}^{\textnormal{forg}}_{A,\mathsf{ATS}}$ of winning an unforgeability and accountability attack game is a negligible function of $\lambda$~\cite{TAPS22}.

\subsection{DTPKE}
A dynamic threshold public-key encryption is a tuple of seven algorithms $(\mathsf{Setup},$ $\mathsf{Join}, \mathsf{Encrypt}, \mathsf{ValidateCT}, \mathsf{ShareDecrypt}, \mathsf{ShareVerify}, \mathsf{Combine})$ invoked as
\begin{gather*}
(mk, ek, dk, vk, ck) \leftarrow \mathsf{KeyGen}(1^{\lambda}),\ (usk, upk, uvk) \leftarrow \mathsf{Join}(mk, id),\\
c\leftarrow \mathsf{Enc}(ek, \mathcal{U}, t', m),\ \{0,1\}\leftarrow \mathsf{ValidateCT}(ek, \mathcal{U}, t', c),\\
\sigma_m^j \leftarrow \mathsf{ShareDecrypt}(dk, pid, usk, c),\ \{0, 1\}\leftarrow \mathsf{ShareVerify}(vk, pid, uvk, c, \sigma_m),\\
\sigma_m\leftarrow \mathsf{Combine}(ck, \mathcal{U}, t', c, \mathcal{N}, \{\sigma^j_m\}_{j\in [t']}).
\end{gather*}

\noindent$\mathsf{Setup}:$ takes as input a security parameter $\lambda$, and outputs the master secret key $mk$, encryption key $ek$, decryption key $dk$, verification key $vk$ and combining key $ck$.

\noindent$\mathsf{Join}:$ takes as input $mk$ and a new user identity $id$, and outputs the private key $usk$, public key $upk$, and verification key $uvk$ of the user.

\noindent$\mathsf{Encrypt}:$ takes as input the encryption key $ek$, the authorized set $\mathcal{U}$, a threshold $t'$, and a message $m$ and outputs a ciphertext $c$.

\noindent$\mathsf{ValidateCT}:$ takes as input $ek, \mathcal{U}, t', c$ and outputs 0 or 1.

\noindent$\mathsf{ShareDecrypt}:$ takes as input $dk, pid, usk, c$ and outputs the decryption share $\sigma_m^j$ or $\bot$.

\noindent$\mathsf{ShareVerify}:$ takes as input $vk, pid, uvk, c, \sigma_m$ and outputs 0 or 1.

\noindent$\mathsf{Combine}:$ takes as input $ck, \mathcal{U}, t', c$, a subset $\mathcal{N}\subset \mathcal{U}$ of $t'$ and a set of $t'$ decryption shares $\{\sigma^j_m\}_{j\in [t']}$, and outputs the plaintext $m$ or $\bot$.

Its non-adaptive adversary, non-adaptive corruption, chosen-plaintext attacks (IND-NAA-NAC-CPA) security is based on the Multi-sequence of Exponents Diffie-Hellman (MSE-DDH) assumption, where $\textbf{Adv}^{\textnormal{ind-cpa}}_{\mathcal{A}, \mathsf{DTPKE}}(l,m,t')\leq \textbf{Adv}^{\textnormal{mse-ddh}}(l, m, t')$~\cite{EUROCRYPT05,Pairing07,DTPKE08}.
We use $\mathsf{Enc}(ek, \mathcal{N}, m)$, $\mathsf{ValidateCT}(ek, \mathcal{N}, c)$, and\\$\mathsf{Combine}(ck, \mathcal{N}, c, \{\sigma^j_m\}_{j\in [t']})$ to simplify the abovementioned $\mathsf{Enc}$, $\mathsf{Combine}$.

\subsection{KASE}
A key-aggregate searchable encryption is a tuple of seven algorithms $(\mathsf{Setup}, \mathsf{Keygen},$ $\mathsf{Encrypt}, \mathsf{Extract}, \mathsf{Trapdoor}, \mathsf{Adjust}, \mathsf{Test})$ invoked as
\begin{gather*}
(\mathcal{B}, \mathcal{PK}, H) \leftarrow \mathsf{KASE.Setup}(\lambda, |G|),\ (mpk, msk) \leftarrow \mathsf{KASE.KeyGen}(\lambda),\\
k_a \leftarrow \mathsf{KASE.Extract}(msk, \mathcal{G}),\ (c^{gid}_1, c^{gid}_2, \{ind_i\}_{i\in \mathcal{N}})\leftarrow \mathsf{KASE.Enc}(mpk, gid, \mathcal{N}),\\
td \leftarrow \mathsf{KASE.Trapdoor}(k_a, pid),\ td^{gid}\leftarrow \mathsf{KASE.Adjust}(\mathcal{B}, \mathcal{PK}, H, gid, \mathcal{G}, td),\\
\{0,1\}\leftarrow \mathsf{KASE.Test}(td^{gid}, (c^{gid}_1, c^{gid}_2, ind)).
\end{gather*}

\noindent$\mathsf{KASE.Setup}:$ takes as input a security parameter $\lambda$ and a size of set $|G|$. It outputs a bilinear map group system $\mathcal{B}=\left\{(p, \mathcal{G}, \mathcal{G}_{1}, e(\cdot, \cdot)\right\}$, the public key $\mathcal{PK} = (g, g_1, \cdots, g_n, g_{n+2}, \cdots,$ $ g_{2n}) \in \mathcal{G}_{2n+1}$ and a one-way hash function $H: \left\{0,1\right\}^{*}\rightarrow\mathcal{G}$.

\noindent$\mathsf{KASE.KeyGen}:$ takes as input $\lambda$, and outputs the key pair $(mpk, msk)$.

\noindent$\mathsf{KASE.Extract}:$ takes as input $msk, \mathcal{G}$, and outputs a aggregate key $k_a$.

\noindent$\mathsf{KASE.Enc}:$ takes as input $mpk$, the file index $gid$ and the keyword set $\mathcal{N}=\left\{pid_i\right\}$, and outputs the ciphertext $(c^{gid}_1, c^{gid}_2, \{ind_i\}_{i\in \mathcal{N}})$.

\noindent$\mathsf{KASE.Trapdoor}:$ takes as input $k_a, pid$ and outputs the trapdoor $td$.

\noindent$\mathsf{KASE.Adjust}:$ takes as input $td$ and outputs the trapdoor $td^{gid}$ for file index $gid$.

\noindent$\mathsf{KASE.Test}:$ takes as input $td^{gid}, (c^{gid}_1, c^{gid}_2, ind)$, and outputs 0 or 1.

KASE achieves controlled searching and query privacy based on the Discrete Logarithm (DL) assumption and the Bilinear Diffie-Hellman Exponent (BDHE) assumption~\cite{EUROCRYPT05}.

\subsection{PKE, COM, SIG}
A public key encryption scheme is a triple of algorithms $(\mathsf{KeyGen}, \mathsf{Encrypt}, \mathsf{Decrypt})$ invoked as
\begin{gather*}
(pk, sk) \leftarrow \mathsf{KeyGen}(1^{\lambda}), c\leftarrow \mathsf{Encrypt}(pk,m),\ m\leftarrow \mathsf{Decrypt}(sk, c).
\end{gather*}

\noindent$\mathsf{KeyGen}:$ takes as input a security parameter $\lambda$, and outputs the key pair $(pk, sk)$.

\noindent$\mathsf{Encrypt}:$ takes as input the public key $pk$ and the message $m$. It outputs the ciphertext $c$.

\noindent$\mathsf{Decrypt}:$ takes as input $m$ and the secret key $sk$, and outputs the plaintext $m$.

A PKE scheme is semantically secure if for every PPT adversary $\mathcal{A}$, $\textbf{Adv}^{\textnormal{ind-cpa}}_{\mathcal{A}, \textnormal{PKE}}(\lambda)$  is negligible~\cite{Katz21}.

A commitment scheme is a pair of algorithms $(\mathsf{Commit}, \mathsf{Verify})$ invoked as
\begin{gather*}
com\leftarrow \mathsf{Commit}(x, r),\ \{0, 1\}\leftarrow \mathsf{Verify}(x, r, com).
\end{gather*}

\noindent$\mathsf{Commit}:$ takes as input a message $x$ and random value $r$ and outputs the commitment $com$.

\noindent$\mathsf{Verify}:$ takes as input $x, r, com$, and outputs 0 or 1.

A COM scheme is secure if it is unconditionally hiding and computationally binding, i.e., for every PPT adversary $\mathcal{A}$, $\textbf{Adv}^{\textnormal{bind}}_{\mathcal{A}, \textnormal{COM}}(\lambda)$ is negligible.

A signature scheme is a triple of algorithms $(\mathsf{KeyGen}, \mathsf{Sign},\mathsf{Verify})$ invoked as
\begin{gather*}
(pk, sk)\leftarrow \mathsf{KeyGen}(1^{\lambda}),\ \sigma\leftarrow \mathsf{Sign}(sk,m),\ \{0, 1\}\leftarrow \mathsf{Verify}(pk, m, \sigma).
\end{gather*}

\noindent$\mathsf{KeyGen}:$ takes as input a security parameter $\lambda$, and outputs the key pair $(pk, sk)$.

\noindent$\mathsf{Encrypt}:$ takes as input the secret key $sk$ and the message $m$. It outputs the signature $\sigma$.

\noindent$\mathsf{Decrypt}:$ takes as input $\sigma,m$ and the public key $pk$, and  outputs 0 or 1.

A SIG scheme is strongly unforgeable if for every PPT adversary $\mathcal{A}$, $\textbf{Adv}^{\textnormal{euf-cma}}_{\mathcal{A}, \textnormal{SIG}}(\lambda)$  is negligible.

\subsection{NIZKP}
A non-interactive zero-knowledge proof protocol enables a prover to convince a verifier that a certain statement is true, without revealing any information about the underlying information for its truth. It involves two algorithms $(\mathsf{P}, \mathsf{V})$ invoked as
\begin{gather*}
	\pi \leftarrow \mathsf{P}(1^{\lambda},m),\ b\leftarrow \mathsf{Verify}(\pi).
\end{gather*}
\noindent$\mathsf{P}:$ takes as input a security parameter $\lambda$ and a relationship $R$, and outputs the proof $\pi$.

\noindent$\mathsf{Verify}:$ takes as input $\pi$. If the verifier rejects, $b=0$, or accepts, $b=1$.

\subsection{Intel SGX2}
Software Guard eXtensions (SGX) is a hardware extension of Intel Architecture that enables an application to establish a protected execution space, i.e., an enclave~\cite{SGX11,SGX12}.
SGX stores enclave pages and SGX structures in the protected memory called Enclave Page Cache (EPC). SGX guarantees confidentiality of code/data and  detection of an integrity violation of an enclave instance from software attacks. SGX allows one to verify that a piece of software has been correctly instantiated on the platform via attestation.
Since SGX imposes limitations regarding memory commitment and reuse of enclave memory, Intel introduces SGX2 to extend the SGX instruction set to include dynamic memory management support for enclaves~\cite{SGX21,SGX22}. SGX2 instructions offer software with more capability to manage memory and page protections from inside an enclave while preserving the security of the SGX architecture and
system software.

\subsection{Consortium Blockchain}
As an underlying technique in Bitcoin, blockchain is a public ledger recording transactions among users who do not fully trust each other in a decentralized network.
The transactions are packed into separate blocks by a set of nodes using a predefined consensus algorithm, and the blocks are sequentially linked into a chain by their cryptographic hashes. These nodes participate in creating new blocks to compete for some rewards such as financial incentives.
Consortium blockchain is a specific blockchain maintained by a group of authorized entities. Only qualified parties are allowed to access the blockchain. It aims to secure transactions between users who do not fully trust each other but work collaboratively toward a common goal. Its consensus process is controlled by the authorized entities.

\subsection{Notations}
To provide further clarification on our scheme, we list important notations used in our paper, which is presented in Table~\ref{tab1}.

\begin{table}[!htb]
\caption{Experimental Parameters}
\begin{center}
\begin{tabular}{|m{1.8cm}<{\centering}|m{4cm}<{\centering}||m{1.8cm}<{\centering}|m{4cm}<{\centering}|} \thickhline
\textbf{Notation} & \textbf{Meaning} & \textbf{Notation} & \textbf{Meaning}\\ \thickhline
		$\lambda$ & security parameter & $gid$ & signer group identifier\\
		$n$ & number of signers & $\mathcal{G}$ & set of signer groups \\
		$n_1$ & number of combiners & $k_a$  & aggregate key  \\
		$n_2$  & number of trances & $\mathcal{N}$ & set of notaries \\
		$n_3$  & number of notaries & $\mathcal{S}$ & signing quorum \\
		$t$ & threshold & $\sigma_i$ & ATS signature\\
		$sk_i$ & key for i-th signer & $\widehat{\sigma_i}$ &  encrypted ATS signature \\
		$\{sk^s_i\}^{n_1}_{i=1}$ & $n_1$ signing keys & $\mathcal{M}$ &  message space\\
		$\{sk^c_i\}^{n_1}_{i=1}$ & $n_1$ combining keys & $m$  &  message for signing \\
		$\{sk^t_i\}^{n_2}_{i=1}$ &  $n_2$ tracing keys & $\sigma$ &  DeTAPS signature \\ \thickhline
\end{tabular}
\end{center}
\label{tab1}
\end{table}

\section{Decentralized, Threshold, dynamically Accountable and Private Signature}
In this section, we formalize the notion of DeTAPS.

\subsection{System Model}
\begin{figure}[h]
\centering
\includegraphics[width=4.8in]{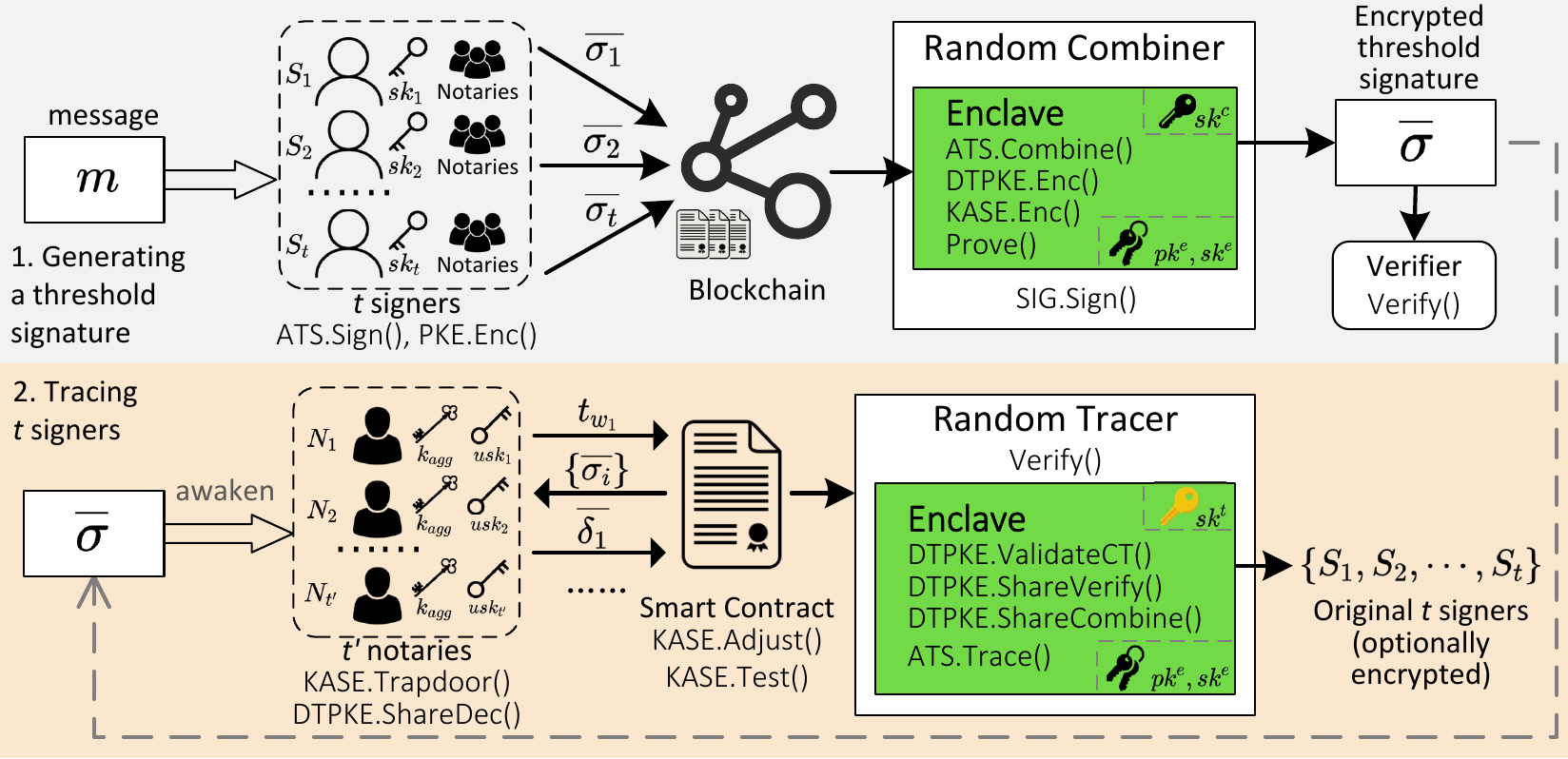}
\caption{System Architecture of DeTAPS.}
 \label{fig1}
\end{figure}

The system architecture of DeTAPS is depicted in Fig.~\ref{fig1}.
It consists of signer, combiner, notary, tracer, and consortium blockchain.

\textbf{Signer.}
When a group of $t$ signers $\mathcal{S} = \{S_1, S_2, \cdots, S_t\}$ prepare to generate a signature on a message $m$, they request the pseudo-identity from $t'$ parties $\mathcal{N} = \{N_1, N_2, \cdots, N_{t'}\}$ as notaries.
Then, each group manager generates a signer group identifier $gid\in \mathcal{G}$ in current signing period and reports it to the CB.
Next, each signer sends a signature share on $m$ to CB.

\textbf{Combiner.}
Each combiner $C_i$ is equipped with a secure enclave $E_i$. $C_i$ has a pair of signing keys and $E_i$ has a pair of encryption keys. The combining key is secured in the $E_i$.
After being elected as a winning node, $C_i$ retrieves all signature shares from the CB and the $E_i$ decrypts them to collect related signature shares and combine them into a complete signature. Next, it generates an encrypted signature via DTPKE, computes an index via KASE, and constructs a non-interactive zero knowledge proof. Finally, $C_i$ signs the message, encrypted signature, index, and the proof.

\textit{Remark 2 (For overlooked signature shares).} During combining in an enclave, there will be overlooked signature shares that exist for the protection of $t$. We do not cast them out of the enclave and retrieve them for the next combining. Instead, we store these shares in the enclave, which has an enough storage space.

\textit{Remark 3 (Why multiple combiners?).} There is only one combiner in TAPS, which is prone to the general problems of centralized model~\cite{BC19}. In DeTAPS, we have distributed such an ability to all blockchain nodes that hold a combining key in a secure enclave. The combining process will be assigned to a randomly node based on the blockchain consensus result. In this way, an adversary will have more difficulty in compromising the actual combiner in current period. This idea also applies to why we have multiple tracers.

\textbf{Notary.}
There is a set of parties working as notaries. In real life, they can be a notary office or a local authority. Each notary $N_i$ has a pseudo-identity, shares an aggregate key, and acts as a user (not necessarily a blockchain node) in the CB network.
Upon a tracing call, each notary $N_i$ computes a trapdoor.
$N_i$ sends the trapdoor to the CB and waits for matching results. If there is a decryption task, $N_i$ verifies the results and then generates a decryption share of the encrypted signature. Next, $N_i$ sends an encrypted response to the CB.

\textbf{Tracer.}
Each tracer $T_j$ is also equipped with a secure enclave $E_j$. The tracing key is secured in the $E_j$.
After being elected as a winning node, $T_j$ retrieves all encrypted decryption shares from the CB and the $E_j$ decrypts them to verify decryption shares.
Finally, $E_j$ collects $t'$ related valid shares to combine a complete signature and trace the original quorum of $t$ tracers.

\textbf{Consortium Blockchain.}
DeTAPS is built upon a decentralized framework where a CB records all the transactions sent by signers, combiners, notaries, and tracers. There are two pools on the CB: a signature share pool $\mathcal{SSL}$ for combiners to track and a decryption share pool $\mathcal{DSL}$ for tracers to monitor. Each of them is deployed on a Smart Contract (SC).

\textbf{Definition 1.} A decentralized, threshold, dynamically accountable and private signature, or DeTAPS, is a tuple of five polynomial time algorithms ${\rm \Pi}= (\mathsf{Setup}, \mathsf{Sign}, \mathsf{Combine},\mathsf{Verify}, \mathsf{Trace})$ as shown in Fig.~\ref{fig2} where
\begin{itemize}
\item[--] $\mathsf{Setup}(1^{\lambda}, n, n_1, n_2, t)\rightarrow (PK, (sk_1, sk_2, \cdots, sk_n), \{sk^c_i\}^{n_1}_{i=1}, \{sk^c_i\}^{n_1}_{i=1}, \{sk^t_j\}^{n_2}_{j=1},$ $\mathcal{G}, k_a)$ is a probabilistic algorithm that takes as input a security parameter $\lambda$, the number of signers $n$, the number of combiners $n_1$, the number of trances $n_2$, and a threshold $t$ to output a public key $PK$, $n$ signer keys $\{sk_1, sk_2, \cdots, sk_n\}$, $n_1$ signing keys $\{sk^s_i\}$, $n_1$ combining keys $\{sk^c_i\}$, $n_2$ tracing keys $\{sk^t_j\}$, a set of signer groups $\mathcal{G}$, and an aggregate key $k_a$.

\item $\mathsf{Sign}(sk_i, m, \mathcal{S}, \mathcal{N}) \rightarrow \widehat{\sigma_i}$ is a probabilistic algorithm run by a signer with a signer key $sk_i$ and a set of notaries $\mathcal{N}$ to generate an encrypted signature share $\widehat{\sigma_i}$ on message $m$ in message space $\mathcal{M}$.

\item $\mathsf{Combine}(sk^c_i, m, \mathcal{S}, \{\widehat{\sigma_j}\}_{j\in \mathcal{S}})\rightarrow \sigma$ a probabilistic algorithm run by a combiner with a combining key $sk^c_i$, a message $m$, a signing quorum $\mathcal{S} = \{S_1, S_2, \cdots, S_t\}$, and $t$ encrypted signature shares $\{\widehat{\sigma_j}\}_{j\in \mathcal{S}}$. If the shares are valid, $\mathsf{Combine}$ outputs a DeTAPS signature $\sigma$.

\item $\mathsf{Verify}(PK, m, \eta) \rightarrow \{0,1\}$ is a deterministic algorithm that verifies the signature $\sigma$ on a message $m$ with respect to the public key $PK$.

\item $\mathsf{Trace}(sk^t_i, m, \sigma) \rightarrow \mathcal{S}$ is a deterministic algorithm run by a tracer with a tracing key $sk^t_i$, a message $m$, and a signature $\sigma$. If $\sigma$ is valid, $\mathsf{Trace}$ outputs a set $\mathcal{S}$ who have generated $\sigma$. Otherwise, it outputs a symbol $\bot$.

\item For \textit{correctness}, we require that for all $t\in [n]$, all $t$-size sets $\mathcal{S}$, all $m\in \mathcal{M}$, and $(PK, (sk_1, sk_2, \cdots, sk_n), \{sk^c_i\}^{n_1}_{i=1}, \{sk^t_i\}^{n_2}_{i=1}, \mathcal{G}, k_a) \leftarrow \mathsf{Setup}(1^{\lambda}, n, t)$ the following two conditions hold:
\begin{equation*}
\begin{aligned}
\textnormal{Pr}[\mathsf{Verify}(PK, m, \mathsf{Combine}(sk^c, sk^s, m, \mathcal{S}, \{\mathsf{Sign}(sk_i, m, \mathcal{S}, \mathcal{N})\}_{i\in \mathcal{S}})) = 1] = 1,\\
\textnormal{Pr}[\mathsf{Trace}(sk^t, m, \mathsf{Combine}(sk^c, sk^s, m, \mathcal{S}, \{\mathsf{Sign}(sk_i, m, \mathcal{S}, \mathcal{N})\}_{i\in \mathcal{S}})) = \mathcal{S}] = 1.
\end{aligned}
\end{equation*}

\end{itemize}

\subsection{Unforgeability and Accountability}
DeTAPS has to satisfy unforgeability and accountability, i.e., existential unforgeability under a chosen message attack with traceability~\cite{TAPS22}.
Informally, unforgeability refers to an adversary that compromises less than $t$ signer cannot generate a valid signature on a message, and
accountability refers to an adversary that compromises $t$ or more signers cannot generate a valid message-signature pair that traces to at least one honest signer.
We formalize these two properties in the adversarial experiment in Fig.~\ref{fig2}. Let $\textbf{Adv}^{\textnormal{forg}}_{\mathcal{A}, {\rm \Pi}}(\lambda)$ be the probability that $\mathcal{A}$ wins the experiment against the DeTAPS scheme ${\rm \Pi}$.

\textbf{Definition 2 (Unforgeability and Accountability).} A DeTAPS scheme ${\rm \Pi}$ is unforgeable and accountable if for all PPT adversaries $A$,
 there is a negligible function $\mathsf{negl}$ such that $\textbf{Adv}^{\textnormal{forg}}_{\mathcal{A}, {\rm \Pi}}(\lambda)\leq \mathsf{negl}(\lambda)$.

\begin{figure}[h]
 \begin{framed}
 \vspace{-0.3cm}
1. $(n, n_1, n_2, t, \mathcal{S}, \mathsf{state}) \overset{\$}{\leftarrow} \mathcal{A}(1^{\lambda})$ where $t\in [n]$, $\mathcal{S}\subseteq [n]$\ \ \ \ \ \ \ \ \ \ \ \ \ \ \ \ \ \ \ \ \ \ \ \ \ \ \ \ \ \ \ \ \ \ \ \ \textbf{Exp$^{\textnormal{forg}}$}\\

\vspace{-0.35cm}
2. $(PK, (sk_1, \cdots, sk_n), \{sk^s_i, sk^c_i\}^{n_1}_{i=1}, \{sk^t_i\}^{n_2}_{i=1}, \mathcal{G}, k_a) \overset{\$}{\leftarrow} \mathsf{Setup}(1^{\lambda}, n, n_1, n_2, t)$\\

\vspace{-0.35cm}
3. $(m', \sigma') \overset{\$}{\leftarrow} \mathcal{A}^{\mathcal{O}(\cdot,\cdot)}(PK, (sk_1, \cdots, sk_n), \{sk^s_i, sk^c_i\}^{n_1}_{i=1}, \{sk^t_j\}^{n_2}_{j=1}, \mathcal{G}, k_a, \mathsf{state})$\\

\vspace{-0.35cm}
where $\mathcal{O}_1(\mathcal{S}_i, m_i)$ returns the signature shares $\{\mathsf{Sign}(sk_j, m_i, \mathcal{S}_i, \mathcal{N})\}_{j\in \mathcal{S}_i}$\\

\vspace{-0.35cm}
Winning condition:\\

\vspace{-0.35cm}
Let $(\mathcal{S}_1, m_1), (\mathcal{S}_2, m_2), \cdots$ be $\mathcal{A}$'s queries to $\mathcal{O}_1$\\

\vspace{-0.35cm}
Let $\mathcal{S} \leftarrow \cup\mathcal{S}_i$, union over all queries to $\mathcal{O}_1(\mathcal{S}_i, m')$, let $\mathcal{S}_t \leftarrow \mathsf{Trace}(sk^t_i, m', \sigma')$\\

\vspace{-0.35cm}
Output 1 if $\mathsf{Verify}(PK, m', \sigma')=1$ and either $\mathcal{S}_t \nsubseteq \mathcal{S}\cup \mathcal{S}'$ or if $\mathcal{S}_t = \mathsf{fail}$
\vspace{-0.3cm}
 \end{framed}
 \centering
\vspace{-0.35cm}
\caption{Experiment of Unforgeability and Accountability.}
  \label{fig2}
\end{figure}

\subsection{Privacy}
(1) Privacy against public. A party who observes a series of $(m,\sigma)$ pairs, acquires nothing about $t$, $t'$ or the signers.
(2) Privacy against signers. Collaborating signers who observe a series  of $(m,\sigma)$ pairs, acquires nothing about $t'$ or signers.
(3) Privacy against combiners. A combiner cannot learn $t$, $t'$, or signers.
(4) Privacy against tracers. A tracer cannot learn $t$, $t'$, or signers.
We formalize the four properties in the adversarial experiment in Fig.~\ref{fig3} and Fig.~\ref{fig4}.

\textbf{Definition 3 (Privacy).}
A DeTAPS scheme is private if for all PPT adversaries $\mathcal{A}$, $\textbf{Adv}^{\textnormal{privP}}_{\mathcal{A}, {\rm \Pi}}(\lambda)$, $\textbf{Adv}^{\textnormal{privS}}_{\mathcal{A}, {\rm \Pi}}(\lambda)$, $\textbf{Adv}^{\textnormal{privC}}_{\mathcal{A}, {\rm \Pi}}(\lambda)$, and $\textbf{Adv}^{\textnormal{privT}}_{\mathcal{A}, {\rm \Pi}}(\lambda)$, are negligible functions of $\lambda$.

\begin{figure*}[t]
\begin{framed}
\vspace{-0.3cm}
1. $b_1\overset{\$}{\leftarrow} \{0,1\}$, $b_2\overset{\$}{\leftarrow} \{0,1\}$\ \ \ \ \ \ \ \ \ \ \ \ \ \ \ \ \ \ \ \ \ \ \ \ \ \ \ \ \ \ \ \ \ \ \ \ \ \ \ \ \ \ \ \ \ \ \ \ \ \ \ \ \ \ \ \ \ \ \ \ \ \ \ \textbf{Exp$^{\textnormal{privP}}$}\\

\vspace{-0.35cm}
2. $(n, n_1, n_2, t_0, t_1, t'_0, t'_1, \mathcal{S}_0, \mathcal{S}_1, \mathcal{N}_0, \mathcal{N}_1, \mathsf{state}) \overset{\$}{\leftarrow} \mathcal{A}(1^{\lambda})$, $t_0, t_1\in [n], t'_0, t'_1\in [n_3]$\\

\vspace{-0.35cm}
3. $(PK, (sk_1, \cdots, sk_n), \{sk^s_i, sk^c_i\}^{n_1}_{i=1}, \{sk^t_j\}^{n_2}_{j=1}, \mathcal{G}, k_a) \overset{\$}{\leftarrow} \mathsf{Setup}(1^{\lambda}, n, n_1, n_2, t_{b_1})$\\

\vspace{-0.35cm}
4. $(b'_1, b'_2) \leftarrow \mathcal{A}^{\mathcal{O}_2(\cdot, \cdot, \cdot), \mathcal{O}_3(\cdot, \cdot, \cdot, \cdot, \cdot), \mathcal{O}_4(\cdot, \cdot)}(PK, \mathsf{state})$\\

\vspace{-0.3cm}
5. Output $(b'_1=b_1)\wedge(b'_2=b_2)$.\\

\vspace{-0.3cm}
where $\mathcal{O}_2(\mathcal{N}_0, \mathcal{N}_1, m||\sigma||gid)$: $\widehat{\sigma}\leftarrow\ \mathsf{PKE.Enc}(pk^e, m||\sigma||\mathcal{N}_{b_2}||gid)$\\

\vspace{-0.3cm}
$\mathcal{O}_3(\mathcal{S}_0, \mathcal{S}_1, \mathcal{N}_0, \mathcal{N}_1, m)$: $\sigma \overset{\$}{\leftarrow} \mathsf{Combine}(sk^c_i, m, \mathcal{S}_{b_1}, \{\mathsf{Sign}(sk_j, m, \mathcal{S}_{b_1}, \mathcal{N}_{b_2}\}_{j})$\\

\vspace{-0.35cm}
\ \ \ \ for $\mathcal{S}_0, \mathcal{S}_1\subseteq [n]$, $|\mathcal{S}_0|=t_0$ and $|\mathcal{S}_1|=t_1$, $\mathcal{N}_0, \mathcal{N}_1\subseteq [n_3]$, $|\mathcal{N}_0|=t'_0$ and $|\mathcal{N}_1|=t'_1$\\

\vspace{-0.35cm}
$\mathcal{O}_4(m, \sigma)$ returns $\mathsf{Trace}(sk^t_i, m, \sigma)$.\\

\vspace{-0.35cm}
\underline{Restriction: if $\sigma$ is computed from $\mathcal{O}_3(\cdot,\cdot, \cdot, \cdot, m)$, $\mathcal{A}$ never queries $\mathcal{O}_4$ at $(m, \sigma)$.\ \ \ \ }\\

\vspace{-0.25cm}
1. $b_1\overset{\$}{\leftarrow} \{0,1\}$, $b_2\overset{\$}{\leftarrow} \{0,1\}$\ \ \ \ \ \ \ \ \ \ \ \ \ \ \ \ \ \ \ \ \ \ \ \ \ \ \ \ \ \ \ \ \ \ \ \ \ \ \ \ \ \ \ \ \ \ \ \ \ \ \ \ \ \ \ \ \ \ \ \ \textbf{Exp$^{\textnormal{privS}}$}\\

\vspace{-0.35cm}
2. $(n, n_1, n_2, t, t'_0, t'_1, \mathcal{S}_0, \mathcal{S}_1, \mathcal{N}_0, \mathcal{N}_1, \mathsf{state}) \overset{\$}{\leftarrow} \mathcal{A}(1^{\lambda})$, $t\in [n], t'_0, t'_1\in [n_3]$\\

\vspace{-0.35cm}
3. $(PK, (sk_1, \cdots, sk_n), \{sk^s_i, sk^c_i\}^{n_1}_{i=1}, \{sk^t_j\}^{n_2}_{j=1}, \mathcal{G}, k_a) \overset{\$}{\leftarrow} \mathsf{Setup}(1^{\lambda}, n, n_1, n_2, t)$\\

\vspace{-0.3cm}
4. $(b'_1, b'_2)\leftarrow \mathcal{A}^{\mathcal{O}_2(\cdot, \cdot, \cdot), \mathcal{O}_4(\cdot, \cdot)}(PK, $\colorbox{LemonChiffon}{$(sk_1, sk_2, \cdots, sk_n)$}, $\mathsf{state})$\\

\vspace{-0.35cm}
5. Output $(b'_1=b_1)\wedge(b'_2=b_2).$\\

\vspace{-0.35cm}
Restriction: $|\mathcal{S}_0|=|\mathcal{S}_1|=t$, $\mathcal{N}_0, \mathcal{N}_1\subseteq [n_3]$, $|\mathcal{N}_0|=t'_0$, $|\mathcal{N}_1|=t'_1$\\
\vspace{-0.45cm}
\end{framed}
\centering
\vspace{-0.35cm}
\caption{Two Experiments of Privacy against the Public and the Signers.}
\label{fig3}
\end{figure*}

\begin{figure*}[h]
\begin{framed}
\vspace{-0.3cm}
1. $b_1\overset{\$}{\leftarrow} \{0,1\}$, $b_2\overset{\$}{\leftarrow} \{0,1\}$\ \ \ \ \ \ \ \ \ \ \ \ \ \ \ \ \ \ \ \ \ \ \ \ \ \ \ \ \ \ \ \ \ \ \ \ \ \ \ \ \ \ \ \ \ \ \ \ \ \ \ \ \ \ \ \ \ \ \ \ \ \ \ \textbf{Exp$^{\textnormal{privC}}$}\\

\vspace{-0.35cm}
2. $(n, n_1, n_2, t_0, t_1, t'_0, t'_1, \mathcal{S}_0, \mathcal{S}_1, \mathcal{N}_0, \mathcal{N}_1, \mathsf{state}) \overset{\$}{\leftarrow} \mathcal{A}(1^{\lambda})$, $t_0, t_1\in [n], t'_0, t'_1\in [n_3]$\\

\vspace{-0.35cm}
3. $(PK, (sk_1, \cdots, sk_n), \{sk^s_i, sk^c_i\}^{n_1}_{i=1}, \{sk^t_j\}^{n_2}_{j=1}, \mathcal{G}, k_a) \overset{\$}{\leftarrow} \mathsf{Setup}(1^{\lambda}, n, n_1, n_2, t_{b_1})$\\

\vspace{-0.25cm}
4. $(b'_1, b'_2) \leftarrow \mathcal{A}^{\mathcal{O}_2(\cdot, \cdot, \cdot), \mathcal{O}_3(\cdot, \cdot, \cdot, \cdot, \cdot), \mathcal{O}_4(\cdot, \cdot)}(PK, $\colorbox{LemonChiffon}{$sk^s_i$}, $\mathsf{state})$, $i\in [n_1]$\\

\vspace{-0.3cm}
5. Output $(b'_1=b_1)\wedge(b'_2=b_2)$.\\

\vspace{-0.35cm}
\underline{Restriction: $sk^s_i$ can be the one used in $\mathcal{O}_3(\mathcal{S}_0, \mathcal{S}_1, \mathcal{N}_0, \mathcal{N}_1, m)$.\ \ \ \ \ \ \ \ \ \ \ \ \ \ \ \ \ \ \ \ \ \ \ \ \ \ \ \ }\\

\vspace{-0.25cm}
1. $b_1\overset{\$}{\leftarrow} \{0,1\}$, $b_2\overset{\$}{\leftarrow} \{0,1\}$\ \ \ \ \ \ \ \ \ \ \ \ \ \ \ \ \ \ \ \ \ \ \ \ \ \ \ \ \ \ \ \ \ \ \ \ \ \ \ \ \ \ \ \ \ \ \ \ \ \ \ \ \ \ \ \ \ \ \ \ \  \textbf{Exp$^{\textnormal{privT}}$}\\

\vspace{-0.35cm}
2. $(n, n_1, n_2, t_0, t_1, t'_0, t'_1, \mathcal{S}_0, \mathcal{S}_1, \mathcal{N}_0, \mathcal{N}_1, \mathsf{state}) \overset{\$}{\leftarrow} \mathcal{A}(1^{\lambda})$, $t_0, t_1\in [n], t'_0, t'_1\in [n_3]$\\

\vspace{-0.35cm}
3. $(PK, (sk_1, \cdots, sk_n), \{sk^s_i, sk^c_i\}^{n_1}_{i=1}, \{sk^t_j\}^{n_2}_{j=1}, \mathcal{G}, k_a) \overset{\$}{\leftarrow} \mathsf{Setup}(1^{\lambda}, n, n_1, n_2, t_{b_1})$\\

\vspace{-0.25cm}
4. $(b'_1, b'_2) \leftarrow \mathcal{A}^{\mathcal{O}_2(\cdot, \cdot, \cdot), \mathcal{O}_3(\cdot, \cdot, \cdot, \cdot, \cdot), \mathcal{O}_4(\cdot, \cdot)}(PK, \mathsf{state})$\\

\vspace{-0.35cm}
5. Output $(b'_1=b_1)\wedge(b'_2=b_2)$.\\

\vspace{-0.45cm}
\end{framed}
\centering
\vspace{-0.35cm}
\caption{Two Experiments of Privacy against the Combiners and the Tracers.}
\label{fig4}
\end{figure*}

\clearpage
\section{A Generic Construction}
In this section, we present a generic construction from a secure ATS scheme. The generic DeTAPS construction consists of seven building blocks:
\begin{itemize}
    \item[--] An ATS$=(\mathsf{KeyGen}, \mathsf{Sign},$ $\mathsf{Combine}, \mathsf{Verify}, \mathsf{Trace})$;
    \item[--] A DTPKE$=(\mathsf{Setup}, \mathsf{Join}, \mathsf{Enc}, \mathsf{Validate}, \mathsf{ShareDecrypt}, \mathsf{ShareVerify},$ $\mathsf{Combine})$;
    \item[--] A KASE$=(\mathsf{Setup}, \mathsf{KeyGen}, \mathsf{Extract}, \mathsf{Enc}, \mathsf{Trapdoor}, \mathsf{Adjust}, \mathsf{Test})$;
    \item[--] A PKE$=(\mathsf{KeyGen}, \mathsf{Encrypt}, \mathsf{Decrypt})$;
    \item[--] A COM$=(\mathsf{Commit}, \mathsf{Verify})$;
    \item[--] A SIG$=(\mathsf{KeyGen}, \mathsf{Sign}, \mathsf{Verify})$;
    \item[--] A non-interactive zero knowledge argument of knowledge $(\mathsf{P}, \mathsf{V})$.
\end{itemize}

The generic DeTAPS scheme is shown in Fig.~\ref{fig5}.
In our construction, a DeTAPS signature on a message $m$ is a tuple $\sigma = (\overline{\sigma}, \pi, \eta)$ where
(1) $\overline{\sigma}$ is a dynamic threshold public-key encryption of an ATS signature $\sigma_m$ on $m$, encrypted by using the ATS public key $pk$,
(2) $\pi$ is a zero-knowledge proof that $\mathcal{N}$ used as notaries is a valid subset of $[n_3]$, the decryption of $\overline{\sigma}$ is a valid ATS signature on $m$, the decryption of $(c_1, c_2, \{ind_i\}_{i\in \mathcal{N}})$ is $(gid, \mathcal{N})$, and
(3) $\eta$ is the combiner's signature on $\overline{\sigma}, \pi)$.

\textit{Remark 4 (Encryption of ATS signature).}
This step initiates dynamically private accountability by involving a quorum of $t'$ notaries $\mathcal{N}$ to encrypt the underlying ATS signature $\sigma_m$. It is triggered by a quorum of $t$ signers who designate $\mathcal{N}$ in generating a signature share $\mathsf{ATS.Sign}(sk_i, m, \mathcal{S})$. When combining $t$ signature shares, an enclave $E_i$ computes a threshold signature $\sigma_m$ and then encrypts $\sigma_m$ by invoking $\overline{\sigma}\leftarrow \mathsf{DTPKE.Enc(ek, \mathcal{N}, \sigma_m)}$. To facilitate successful tracing, each relevant notary $N_j$ has to generate a decryption share of $\overline{\sigma}$ by using $\sigma^j_m \leftarrow \mathsf{DTPKE.ShareDecrypt}(dk, pid_j, usk_j, \overline{\sigma})$ for a tracer to combine $t'$ decryption shares and run $\mathcal{S}\leftarrow \mathsf{ATS.Trace}(pk, m, \sigma_m)$.

\textit{Remark 5 (Encryption of $gid$ and $\mathcal{N}$).}
After an encrypted threshold signature is published and its signers are held accountable, we need to awaken its notaries to decrypt the encrypted threshold signature. To this end, we resort to KASE. The enclave creates a index by computing $(c^{gid}_1, c^{gid}_2, \{ind_i\}_{i\in \mathcal{N}})\leftarrow \mathsf{KASE.Enc}(mpk, gid, \mathcal{N})$ where $gid$ resembles file index and items in $\mathcal{N}=\{pid_i\}$ are keywords. Since the $|\{ind_i\}| = t'$, we use some dummy $pid$s to hide $t'$.
In tracing, a notary uses an aggregate-key $k_a$ to compute a trapdoor $td_i\leftarrow \mathsf{KASE.Trapdoor}(k_a, pid_i)$ for the SC to look for matching indexes.

\textit{Remark 6 (The generation of $\pi$).} There are five parts in $\pi$. The first one and second one are done by committing to a vector, proving that every commitment is well formed~\cite{TAPS22,PKC98}, and generating NIZKPs using the Fiat-Shamir transform.
The last three are done by generating NIZKPs as well.

\textit{Remark 7 (Protect $t$ from Tracer).}
After the original quorum of signers is revealed in an enclave, we can encrypt their identities by a target party's public key for directional tracing.
If not, we can just wait for some time to reveal a batch of quorums including $t$ and then re-setup the system with a new $t$.

\textit{Correctness.} DeTAPS is correct if the $\mathsf{ATS}$ scheme, $\mathsf{DTPKE}$ scheme, $\mathsf{KASE}$ scheme, $\mathsf{PKE}$ scheme, $\mathsf{COM}$ scheme, $\mathsf{SIG}$ scheme, and $(\mathsf{P}, \mathsf{V})$ are correct.

\begin{figure}[H]
 \begin{framed}
\vspace{-0.2cm}
\underline{$\mathsf{Setup}(1^{\lambda}, n, n_1, n_2, t):$}\\

\vspace{-0.33cm}
$1.\ (pk, \{sk_i\}^{n}_{i=1}) \leftarrow \mathsf{ATS.KeyGen}(1^{\lambda}, n, t)$\ \textcolor{darkgrey}{//$n$: number of signers}\\

\vspace{-0.33cm}
$2.\ r_{pk} \leftarrow \mathcal{R}_{\lambda}$, $\textsf{com}_{pk} \leftarrow \mathsf{COM.Comm}(pk, r_{pk})$\\

\vspace{-0.33cm}
$3.\ (mk, ek, dk, vk, ck) \leftarrow \mathsf{DTPKE.Setup}(1^{\lambda})$\\

\vspace{-0.33cm}
$4.\ (pk^s_j, sk^s_j) \leftarrow \mathsf{SIG.KeyGen}(1^{\lambda}, j)$, $j\in [n_1]$\ \ \ \textcolor{darkgrey}{//Combiner $C_j$'s signing keys}\\

\vspace{-0.33cm}
$5.\ (pk^e_j, sk^e_j) \leftarrow \mathsf{PKE.KeyGen}(1^{\lambda}, j)$, $j\in [n_1]$\ \textcolor{darkgrey}{//Enclave $E_j$'s encryption keys}\\

\vspace{-0.33cm}
$6.\ (usk_o, upk_o, uvk_o) \leftarrow \mathsf{DTPKE.Join}(mk, o)$, $o\in [n_3]$\ \textcolor{darkgrey}{//$n_3$: number of notaries}\\

\vspace{-0.33cm}
$7.\ (\mathcal{B}, \mathcal{PK}, H) \leftarrow \mathsf{KASE.Setup}(\lambda, |G|)$\\

\vspace{-0.33cm}
$8.\ (mpk, msk) \leftarrow \mathsf{KASE.KeyGen}(\lambda)$\\

\vspace{-0.33cm}
$9.\ k_a \leftarrow \mathsf{KASE.Extract}(msk, \mathcal{G})$\\

\vspace{-0.33cm}
$10.\ sk_j^c \leftarrow (\tikzmark{id}{\colorbox{grey}{$pk, sk^e_j, t, ek, r_{pk}$}}), \tikzmark{kd}{$j\in [n_1]$}$\ \textcolor{darkgrey}{//Combining key}\\
\begin{tikzpicture}[overlay, remember picture,node distance =0.10cm]
    \node [darkgrey](cd) [above right=of kd ]{\textbf{\underline{stored in Enclave $E_j$}}};
    \draw[darkgrey,->,thick] (cd) to [in=10,out=-165] (id);
   \end{tikzpicture}

\vspace{-0.33cm}
$11.\ sk_j^t \leftarrow ($\colorbox{grey}{$sk^e_j, ck, pk$}), $j\in [n_2]$\ \ \ \ \ \ \ \ \ \ \ \ \textcolor{darkgrey}{//Tracing key}\\

\vspace{-0.33cm}
$12.\ gid \leftarrow \mathsf{HASH}(GID, time)$, $GID\in \mathcal{G}$\ \ \ \textcolor{darkgrey}{//$gid$: signer group identifier}\\

\vspace{-0.33cm}
$13.\ PK \leftarrow (\textsf{com}_{pk}, ek, dk, vk, \{pk^s_i\}^{n_j}_{j=1}, \{pk^e_j\}^{n_1}_{j=1}, \mathcal{B}, \mathcal{PK}, H, mpk, \{gid\})$\\

\vspace{-0.33cm}
$14.$ Output $(PK, \{sk_i\}^{n}_{i=1}, \{sk^s_j\}^{n_1}_{j=1}, \{sk^c_j\}^{n_1}_{j=1}, \{sk^t_j\}^{n_2}_{j=1}, \mathcal{G}, k_a)$\\

\vspace{-0.33cm}
\underline{$\mathsf{Sign}(sk_i, m, \mathcal{S}, \mathcal{N}) \rightarrow \widehat{\sigma_i}:$}\\

\vspace{-0.33cm}
$1.\ \sigma_i\leftarrow \textsf{ATS.Sign}(sk_i, m, \mathcal{S})$\\

\vspace{-0.33cm}
$2.\ \widehat{\sigma}_i\leftarrow \mathsf{PKE.Enc}(pk^e_j, m||\sigma_i||\mathcal{N}||gid)$\ \textcolor{darkgrey}{//Sent to CB to be processed by a combiner}\\

\vspace{-0.33cm}
\underline{$\mathsf{Combine}(\colorbox{grey}{$sk^c_j=(pk, sk^e_j, t, ek, r_{pk}$)}, sk^s_j, m, \mathcal{S}, \{\widehat{\sigma_i}\}_{i\in \mathcal{S}})\rightarrow \sigma$}\\

\colorbox{grey}{
$1.\ (m||\sigma_i||\mathcal{N}||gid)\leftarrow \mathsf{PKE.Dec}(\widehat{\sigma}_i, sk^e_j)$\ \textcolor{darkgrey}{//\textbf{Enter $E_i$} here; $\sigma_i\in$ Tx pool}}\\

\vspace{-0.33cm}
\colorbox{grey}{
$2.\ \sigma^m \leftarrow \mathsf{ATS.Combine}(pk, m, \mathcal{S}, \{\sigma_i\}_{i\in \mathcal{S}})$}\\

\vspace{-0.33cm}
\colorbox{grey}{
$3.\ \overline{\sigma}\leftarrow \mathsf{DTPKE.Enc}(ek, \mathcal{N}, \sigma^m)$}\\

\vspace{-0.33cm}
\colorbox{grey}{
$4.\ (c^{gid}_1, c^{gid}_2, \{ind_o\}_{o\in \mathcal{N}})\leftarrow \mathsf{KASE.Enc}(mpk, gid, \mathcal{N})$}\\

\vspace{-0.33cm}
\colorbox{grey}{
5. Generate a proof for the relation:}
\vspace{-0.23cm}
\begin{equation*}
\mathcal{R}((t', \textsf{com}_{pk}, ek, mpk, m, \overline{\sigma}, gid, c^{gid}_1, c^{gid}_2, \{ind_o\}_{o\in \mathcal{N}}); (\mathcal{N}, \sigma^m, r_{pk}, pk)) = \textnormal{1 iff}
\vspace{-0.23cm}
\end{equation*}
\begin{equation*}
      \left\{
      \begin{aligned}
      &\mathcal{N} \subseteq [n_3]\\
      &\mathsf{ATS.Verify}(pk, m, \sigma^m) = 1,\ \mathsf{COM.Verify}(pk, r_{pk}, \textsf{com}_{pk}) = 1,\\
      &\overline{\sigma} = \mathsf{DTPKE.Enc}(ek, \mathcal{N}, \sigma^m),\ (c_1, c_2, \{ind_o\})\leftarrow \mathsf{KASE.Enc}(mpk, gid, \mathcal{N}).\\
      \end{aligned}
      \right\}
\end{equation*}

\vspace{-0.15cm}
$6.\ \eta\leftarrow \mathsf{SIG.Sign}(sk^s_j, (m, \overline{\sigma}, c^{gid}_1, c^{gid}_2, \{ind_o\}_{o\in \mathcal{N}}, \pi))$\\

\vspace{-0.33cm}
$7.$ Output a DeTAPS signature $\sigma\leftarrow (\overline{\sigma}, c^{gid}_1, c^{gid}_2, \{ind_o\}_{o\in \mathcal{N}}, \pi, \eta)$\\

\vspace{-0.33cm}
\underline{$\mathsf{Verify}(PK, m, \sigma = (\overline{\sigma}, \pi, \eta))\rightarrow\ \{0,1\}$}\\

\vspace{-0.33cm}
1. Accept $\sigma$ if $\mathsf{SIG.Verify}(pk^s_j, m, \sigma) = 1$ and $\pi$ is valid; reject otherwise.\\

\vspace{-0.33cm}
\underline{$\mathsf{Trace}(sk^t_j = (\colorbox{grey}{$sk^e_j, ck, pk$}), m, \sigma = ( \overline{\sigma}, \pi, \eta)) \rightarrow \mathcal{S}$}\\
$1.\ td_o \leftarrow \mathsf{KASE.Trapdoor}(k_a, pid_o)$\ \textcolor{darkgrey}{//For each notary with a pseudo identity $pid_o$}\\

\vspace{-0.33cm}
$2.\ td^{gid}_o\leftarrow \mathsf{KASE.Adjust}(\mathcal{B}, \mathcal{PK}, H, gid, \mathcal{G}, td_o)$\ \textcolor{darkgrey}{//For each $gid$ and $td_o$}\\

\vspace{-0.33cm}
$3.\ \{0,1\}\leftarrow \mathsf{KASE.Test}(td^{gid}_o, (c^{gid}_1, c^{gid}_2, \{ind_o\}))$\ \textcolor{darkgrey}{//Search all indexes to locate $\overline{\sigma}$}\\

\vspace{-0.33cm}
4. If $\mathsf{SIG.Verify}(pk^s_j, (m, \overline{\sigma}, \pi), \eta) \neq 1$, output fail and return.\\

\vspace{-0.33cm}
5. $\sigma^m_o \leftarrow \mathsf{DTPKE.ShareDecrypt}(dk, pid_o, usk_o, \overline{\sigma})$\ \textcolor{darkgrey}{//For each $pid_o$ and each $m$}\\

\vspace{-0.33cm}
6. $\overline{uvk_o}\leftarrow \mathsf{PKE.Enc}(pid_o, uvk_o, \delta_o, pk^e_j)$\\

\vspace{-0.33cm}
\colorbox{grey}{
7. $(pid_o, uvk_o, \delta_o) \leftarrow \mathsf{PKE.Dec}(\overline{uvk_o}, sk^e_j)$}\\

\vspace{-0.33cm}
\colorbox{grey}{
8. $\{0,1\}\leftarrow \mathsf{DTPKE.ValidateCT}(ek, \mathcal{N}, \overline{\sigma})$}\ \textcolor{darkgrey}{//Assume that $\overline{\sigma}$ is preloaded}\\

\vspace{-0.3cm}
\colorbox{grey}{
9. $\{0, 1\}\leftarrow \mathsf{DTPKE.ShareVerify}(vk, pid_o, uvk_o, \overline{\sigma}, \delta_o)$}\\

\vspace{-0.3cm}
\colorbox{grey}{
10. $\sigma^m\leftarrow \mathsf{DTPKE.Combine}(ck, \mathcal{N}, \overline{\sigma}, \{\delta_o\}_{o\in \mathcal{N}})$}\\

\vspace{-0.33cm}
\colorbox{grey}{
11. $\mathcal{S}\leftarrow \mathsf{ATS.Trace}(pk, m, \sigma^m)$.}
\vspace{-0.2cm}
 \end{framed}
 \centering
 \caption{The generic DeTAPS scheme}
 \label{fig5}
\end{figure}

\section{Security and Privacy of DeTAPS}
Now we prove that the generic scheme is secure, accountable, and private.

\textbf{Theorem 1.} The generic DeTAPS scheme ${\rm \Pi}$ in Fig.~\ref{fig5} is unforgeable, accountable, and private, assuming that the underlying $\mathsf{ATS}$ is secure, the $\mathsf{DTPKE}$ is IND-NAA-NAC-CPA secure, the $\mathsf{KASE}$ is privacy-preserving, the $\mathsf{PKE}$ is semantically secure, the $(\mathsf{P}, \mathsf{V})$ is an argument of knowledge and honest verifier zero knowledge (HVZK), the $\mathsf{COM}$ is hiding and binding, and the $\mathsf{SIG}$ is strongly unforgeable.

The proof of Theorem 1 is captured in the following five lemmas.

$\textbf{Lemma 1}$. The generic DeTAPS scheme ${\rm \Pi}$ is unforgeable and accountable if the ATS is secure, the $(\mathsf{P}, \mathsf{V})$ is an argument of knowledge, and $\mathsf{COM}$ is blinding, i.e., for all PPT adversaries $\mathcal{A}$, $\mathcal{A}_1$, and $\mathcal{A}_2$, such that
\begin{equation}
\textnormal{\textbf{Adv}}^{\textnormal{forg}}_{\mathcal{A}, {\rm \Pi}}(\lambda) \leq \left(\textnormal{\textbf{Adv}}^{\textnormal{forg}}_{\mathcal{A}_1, \mathsf{ATS}}(\lambda) + \textnormal{\textbf{Adv}}^{\textnormal{bind}}_{\mathcal{A}_2, \mathsf{COM}}(\lambda)\right) \cdot \alpha(\lambda) + \beta(\lambda),
\end{equation}
where $\alpha$ and $\beta$ are the knowledge error and tightness of the proof system.

\textit{Proof.} We prove Lemma 1 by defining experiments Exp 0, Exp 1, and Exp 2.

\textbf{Exp 0.} It is the experiment of unforgeability and accountability \textbf{Exp$^{\textnormal{forg}}$} defined in Fig.~\ref{fig2} applied to ${\rm \Pi}$. If $E_0$ stands for $\mathcal{A}$ wins $Exp_0$, then
\begin{equation}
\textnormal{\textbf{Adv}}^{\textnormal{forg}}_{\mathcal{A}, {\rm \Pi}}(\lambda)=\textnormal{Pr}[E_0].
\end{equation}

\textbf{Exp 1.} It is identical to Exp 0 with a strengthened winning condition: the adversary has to output a valid forgery $(m', \sigma')$ where $\sigma' = (\overline{\sigma}', c^{gid'}_1, c^{gid'}_2, \{ind'_o\},$ $\pi', \eta')$ with a witness satisfying $\mathcal{R}((t', \textsf{com}_{pk}, ek, mpk, m', \overline{\sigma}', gid', c^{gid'}_1, c^{gid'}_2,$ $\{ind'_o\});$ $(\mathcal{N}'', \sigma''^m, r''_{pk}, pk'')) = 1$.

Assume $\mathcal{A}'$ is an adversary in Exp 1. It invokes $\mathcal{A}$ and answers to $\mathcal{A}$'s queries until receives from $\mathcal{A}$ the $(m', \overline{\sigma}', c^{gid'}_1, c^{gid'}_2, \{ind'_o\})$ to provide a statement $(t', \textsf{com}_{pk}, ek, mpk, m', \overline{\sigma}', gid', c^{gid'}_1, c^{gid'}_2, \{ind'_o\})$.
$\mathcal{A}'$ executes the extractor $Ext$ for $(\mathsf{P}, \mathsf{V})$ on $\mathcal{A}$'s remaining execution.
$Ext$ produces a witness $w=(\mathcal{N}'', \sigma''^m, r''_{pk}, pk'')$. $\mathcal{A}'$ uses $w$ and $sk^s$ to generate $\pi'$ and $\eta'$ such that $\sigma' = (\overline{\sigma}', c^{gid'}_1, c^{gid'}_2, \{ind'_o\}, \pi', \eta')$ is a valid signature on $m'$. $\mathcal{A}'$ outputs $(m', \sigma')$ and $w$.
By definition of $Ext$, if $E_1$ stands for $\mathcal{A}'$ wins Exp 1, then
\begin{equation}
\textnormal{Pr}[E_1] \geq (\textnormal{Pr}[E_0] - \alpha(\lambda))/\beta(\lambda).
\end{equation}

\textbf{Exp 2.} The adversary now has $pk$ and $r_{pk}$. We strengthen the winning condition by requiring $pk = pk''$. Let $E_2$ stand for $\mathcal{A}$ wins Exp 2 and $E$ stand for $pk\neq pk''$. Therefore, $\textnormal{Pr}[E_2] = \textnormal{Pr}[E_1\wedge \neg E] \geq \textnormal{Pr}[E_1] - \textnormal{Pr}[E]$. Assume that there is an adversary $\mathcal{A}_2$ such that $\textnormal{Pr}[E] = \textnormal{\textbf{Adv}}^{\textnormal{bind}}_{\mathcal{A}_2, \mathsf{COM}}(\lambda)$. We have
\begin{equation}
\textnormal{Pr}[E_2] \geq \textnormal{Pr}[E_1] - \textnormal{\textbf{Adv}}^{\textnormal{bind}}_{\mathcal{A}_2, \mathsf{COM}}(\lambda).
\end{equation}

Next, we construct an adversary $\mathcal{A}_1$ that invokes $\mathcal{A}$  and answers to $\mathcal{A}$'s queries. When $\mathcal{A}$ outputs a forgery $(m', \sigma')$ and a witness $(\mathcal{N}'', \sigma''^m, r''_{pk}, pk'')$ that meet the winning condition of Exp 1 and Exp 2, $\mathcal{A}_1$ outputs $(m', \sigma''^m)$.
By $\mathcal{R}$, we have $\sigma''^m$ is a valid signature on $m'$ with respect to $pk''$. By Exp 2, we have $pk=pk''$. Therefore, if $\mathcal{A}$ wins Exp 2, then $(m', \sigma''^m)$ is a valid forgery for the $\mathsf{ATS}$ scheme. Since the $\mathsf{ATS}$ is secure, we have that $\textnormal{Pr}[E_2]$ is at most negligible, i.e.,
\begin{equation}
\textnormal{\textbf{Adv}}^{\textnormal{forg}}_{\mathcal{A}_1, \mathsf{ATS}}(\lambda) \geq \textnormal{Pr}[E_2].
\end{equation}

Finally, combining (2), (3), (4), and (5) proves (1). This completes the proof of the lemma.\hfill $\Box$

$\textbf{Lemma 2}$. The generic DeTAPS scheme ${\rm \Pi}$ is private against the public if
the $\mathsf{PKE}$ is semantically secure,
the $\mathsf{SIG}$ is strongly unforgeable,
the $(\mathsf{P}, \mathsf{V})$ is HVZK,
the $\mathsf{KASE}$ is privacy-preserving,
the $\mathsf{DTPKE}$ is IND-NAA-NAC-CPA secure,
the $\mathsf{COM}$ is hiding,
i.e., for all PPT adversaries $\mathcal{A}$, there exists adversaries $\mathcal{A}_1$, $\mathcal{A}_2$, $\mathcal{A}_3$, $\mathcal{A}_4$, $\mathcal{A}_5$, and $\mathcal{A}_6$, such that
\begin{equation}
\begin{aligned}
\textnormal{\textbf{Adv}}^{\textnormal{priP}}_{\mathcal{A}, {\rm \Pi}}(\lambda) \leq 2\left(
\textnormal{\textbf{Adv}}^{\textnormal{indcpa}}_{\mathcal{A}_1, \mathsf{PKE}}(\lambda) +
\textnormal{\textbf{Adv}}^{\textnormal{euf-cma}}_{\mathcal{A}_2, \mathsf{SIG}}(\lambda) + Q\cdot \textnormal{\textbf{Adv}}^{\textnormal{hvzk}}_{\mathcal{A}_3, (\mathsf{P},\mathsf{V})}(\lambda) +\right.\\
\left.\epsilon_{\mathcal{A}_4}(\lambda) +
\textnormal{\textbf{Adv}}^{\textnormal{indcka}}_{\mathcal{A}_5, \mathsf{KASE}}(\lambda) +
\textnormal{\textbf{Adv}}^{\textnormal{ind-cpa}}_{\mathcal{A}_6, \mathsf{DTPKE}}(\lambda) +\right)
\end{aligned}
\end{equation}
where $\epsilon(\lambda)_{\mathcal{A}_4}$ is hiding statistical distance of $\mathsf{COM}$ and $Q$ is query number.

\textit{Proof.} We prove Lemma 2 by defining seven experiments.

\textbf{Exp 0.} It is the experiment of privacy against the public \textbf{Exp$^{\textnormal{priP}}$} defined in Fig.~\ref{fig3} applied to ${\rm \Pi}$. If $E_0$ stands for $\mathcal{A}$ wins Exp 0, then
\begin{equation}
\textnormal{\textbf{Adv}}^{\textnormal{priP}}_{\mathcal{A}, {\rm \Pi}}(\lambda) = |2\textnormal{Pr}[E_0]-1|.
\end{equation}

\textbf{Exp 1.} It is identical to Exp 0 except that the signing oracle $\mathcal{O}_1(\mathcal{S}_0, \mathcal{S}_1, \mathcal{N}_0, \mathcal{N}_1,$ $m)$ is modified such that step 2 of $\mathsf{Sign}$ in Fig.~\ref{fig5} now returns $\widehat{\sigma}_i\leftarrow \mathsf{PKE.Enc}(pk^e_j, 0)$, where 0 is encrypted instead of $(m||\sigma_i||\mathcal{N}||gid)$.
Since $\mathsf{PKE}$ is semantically secure, $\mathcal{A}_1$'s $\textnormal{\textbf{Adv}}$ in Exp 1 is indistinguishable from its $\textnormal{\textbf{Adv}}$ in Exp 0, i.e., say $E_1$ stands for $\mathcal{A}_1$ wins Exp 1,
\begin{equation}
|\textnormal{Pr}[E_1]-\textnormal{Pr}[E_0]| \leq \textnormal{\textbf{Adv}}^{\textnormal{ind-cpa}}_{\mathcal{A}_1, \mathsf{PKE}}(\lambda).
\end{equation}

\textbf{Exp 2.} It is identical to Exp 0 except that responses to $\mathcal{O}_2(m, \sigma)$ are $\mathsf{fail}$. If $\mathsf{SIG}$ is strongly unforgeable, $\mathcal{A}_1$'s $\textnormal{\textbf{Adv}}$ in Exp 2 is indistinguishable from its $\textnormal{\textbf{Adv}}$ in Exp 1, i.e., say $E_2$ stands for $\mathcal{A}_2$ wins Exp 2,
\begin{equation}
|\textnormal{Pr}[E_2]-\textnormal{Pr}[E_1]| \leq \textnormal{\textbf{Adv}}^{\textnormal{euf-cma}}_{\mathcal{A}_2, \mathsf{SIG}}(\lambda).
\end{equation}

\textbf{Exp 3.} It is identical to Exp 2 except that the signing oracle $\mathcal{O}_1(\mathcal{S}_0, \mathcal{S}_1, \mathcal{N}_0, \mathcal{N}_1,$ $m)$ is modified such that step 5 of $\mathsf{Combine}$ now generates a proof $\pi$ by using the simulator, which is given $(t', \textsf{com}_{pk}, ek, mpk, m, \overline{\sigma}, gid, c^{gid}_1, c^{gid}_2, \{ind_o\}_{o\in \mathcal{N}})$ as input.
Since the simulated proofs are computationally indistinguishable from real proofs, $\mathcal{A}_3$'s $\textnormal{\textbf{Adv}}$ in Exp 3 is indistinguishable from its $\textnormal{\textbf{Adv}}$ in Exp 2, i.e., say $E_3$ stands for $\mathcal{A}_2$ wins Exp 3,
\begin{equation}
|\textnormal{Pr}[E_3]-\textnormal{Pr}[E_2]| \leq Q\cdot \textnormal{\textbf{Adv}}^{\textnormal{hvzk}}_{\mathcal{A}_3, (\mathsf{P}, \mathsf{V})}(\lambda).
\end{equation}

\textbf{Exp 4.} It is identical to Exp 3 except that step 2 of $\mathsf{Setup}$  in Fig.~\ref{fig5} is modified such that $r_{pk} \leftarrow \mathcal{R}_{\lambda}$, $\textsf{com}_{pk} \leftarrow \mathsf{COM.Comm}(0, r_{pk})$, where 0 is committed instead of $pk$.
Since $\mathsf{COM}$ is hiding, the adversary's $\textnormal{\textbf{Adv}}$ in Exp 4 is indistinguishable from its $\textnormal{\textbf{Adv}}$ in Exp 3, i.e., say $E_3$ stands for $\mathcal{A}_4$ wins Exp 4,
\begin{equation}
|\textnormal{Pr}[E_4]-\textnormal{Pr}[E_3]| \leq \epsilon_{\mathcal{A}_4}(\lambda).
\end{equation}

\textbf{Exp 5.} It is identical to Exp 4 except that the signing oracle $\mathcal{O}_1(\mathcal{S}_0, \mathcal{S}_1, \mathcal{N}_0, \mathcal{N}_1, m)$ is modified such that step 4 of $\mathsf{Combine}$ now returns $(c^{gid}_1, c^{gid}_2, \{ind_o\}_{o\in \mathcal{N}})\leftarrow \mathsf{KASE.Enc}(mpk, gid, \{r_j\})$, where a random set is encrypted instead of $\mathcal{N}$.
Since $\mathsf{KASE}$ is privacy-preserving, $\mathcal{A}_4$'s $\textnormal{\textbf{Adv}}$ in Exp 5 is indistinguishable from its $\textnormal{\textbf{Adv}}$ in Exp 4, i.e., say $E_5$ stands for $\mathcal{A}_5$ wins Exp 5,
\begin{equation}
|\textnormal{Pr}[E_5]-\textnormal{Pr}[E_4]| \leq \textnormal{\textbf{Adv}}^{\textnormal{indcka}}_{\mathcal{A}_5, \mathsf{KASE}}(\lambda).
\end{equation}

\textbf{Exp 6.} It is identical to Exp 5 except that the signing oracle $\mathcal{O}_1(\mathcal{S}_0, \mathcal{S}_1, \mathcal{N}_0, \mathcal{N}_1,$ $m)$ is modified such that step 3 of $\mathsf{Combine}$ now returns $\overline{\sigma}\leftarrow \mathsf{DTPKE.Enc}(ek,$ $\mathcal{N}, 0)$, where 0 is encrypted instead of $\sigma^m$.
Since $\mathsf{DTPKE}$ is secure, $\mathcal{A}_6$'s $\textnormal{\textbf{Adv}}$ in Exp 6 is indistinguishable from its $\textnormal{\textbf{Adv}}$ in Exp 5, i.e., say $E_5$ stands for $\mathcal{A}_6$ wins Exp 6,
\begin{equation}
|\textnormal{Pr}[E_6]-\textnormal{Pr}[E_5]| \leq \textnormal{\textbf{Adv}}^{\textnormal{ind-cpa}}_{\mathcal{A}_6, \mathsf{DTPKE}}(\lambda).
\end{equation}

In Exp 6, $\mathcal{A}_6$'s view is independent of $b$. Consequently, $\mathcal{A}_6$ has no advantage in Exp 6, i.e.,
\begin{equation}
\textnormal{Pr}[E_6] = 1/2.
\end{equation}
Lastly, combining (7)-(14) proves (6). This completes the proof of lemma 2.\hfill $\Box$

$\textbf{Lemma 3}$. The generic DeTAPS scheme ${\rm \Pi}$ is private against the signers
\textit{Proof.} The proof of Lemma 3 is identical to the proof of Lemma 2.\hfill $\Box$

$\textbf{Lemma 4}$. The generic DeTAPS scheme ${\rm \Pi}$ is private against the combiners
\textit{Proof.} The proof of Lemma 4 is almost identical to the proof of Lemma 2 except that the Exp 1 is removed because the combiner has the signing key, i.e.,
\begin{equation}
\begin{aligned}
&\textnormal{\textbf{Adv}}^{\textnormal{priP}}_{\mathcal{A}, {\rm \Pi}}(\lambda) \leq 2\left(
Q\cdot \textnormal{\textbf{Adv}}^{\textnormal{hvzk}}_{\mathcal{A}_1, (\mathsf{P},\mathsf{Verify})}(\lambda) + \epsilon(\lambda) + \right.\\
&\left.\textnormal{\textbf{Adv}}^{\textnormal{indcka}}_{\mathcal{A}_3, \mathsf{KASE}}(\lambda)+
\textnormal{\textbf{Adv}}^{\textnormal{ind-cpa}}_{\mathcal{A}_4, \mathsf{DTPKE}}(\lambda) +
\textnormal{\textbf{Adv}}^{\textnormal{indcpa}}_{\mathcal{A}_5, \mathsf{PKE}}(\lambda)\right)
\end{aligned}
\end{equation}

Similarly, combining (7), (9)-(14) proves (15). This completes the proof of Lemma 4.\hfill $\Box$

$\textbf{Lemma 5}$. The generic DeTAPS scheme ${\rm \Pi}$ is private against the tracers.
\textit{Proof.} Although the tracer carries out the tracing process within its enclave, its view is the same as one from the public. Therefore, the proof of Lemma 5 is identical to the proof of Lemma 2.\hfill $\Box$

\section{Performance Evaluation}
In this section, we build a prototype of DeTAPS based on Intel SGX2 and Ethereum blockchain.
We evaluate its performance regarding computational costs and communication overhead of five phases.

\subsection{Experimental Settings}
\textbf{Dataset and Parameters}.
Since there are no specialized datasets, we synthesize the input data.
Table~\ref{tab2} lists key experimental paraments.
See codes on \uline{github.com/UbiPLab/DeTAPS} and full version on \uline{arxiv.org/abs/2304.07937}.

\vspace{-0.5cm}
\begin{table}[!htb]
\caption{Experimental Parameters}
\begin{center}
\begin{tabular}{|m{2.4cm}<{\centering}|m{3cm}<{\centering}||m{2.4cm}<{\centering}|m{2.4cm}<{\centering}|} \thickhline
\textbf{Parameter} & \textbf{Value} & \textbf{Parameter} & \textbf{Value}\\ \thickhline
$n$, $n_3$ & $[10, 50]$, $[10, 50]$                & $|m|$, $\lambda$ & $[1, 10]$, $512$  \\ \hline
$n_1$, $n_2$, $n_4$ & 5, 5, $[100, 1000]$  & $t$, $t'$ & \{5, 10, 15\}   \\ \hline
\end{tabular}
\end{center}
\label{tab2}
\end{table}

\begin{figure}[h]
\centering
\includegraphics[width=4.1in]{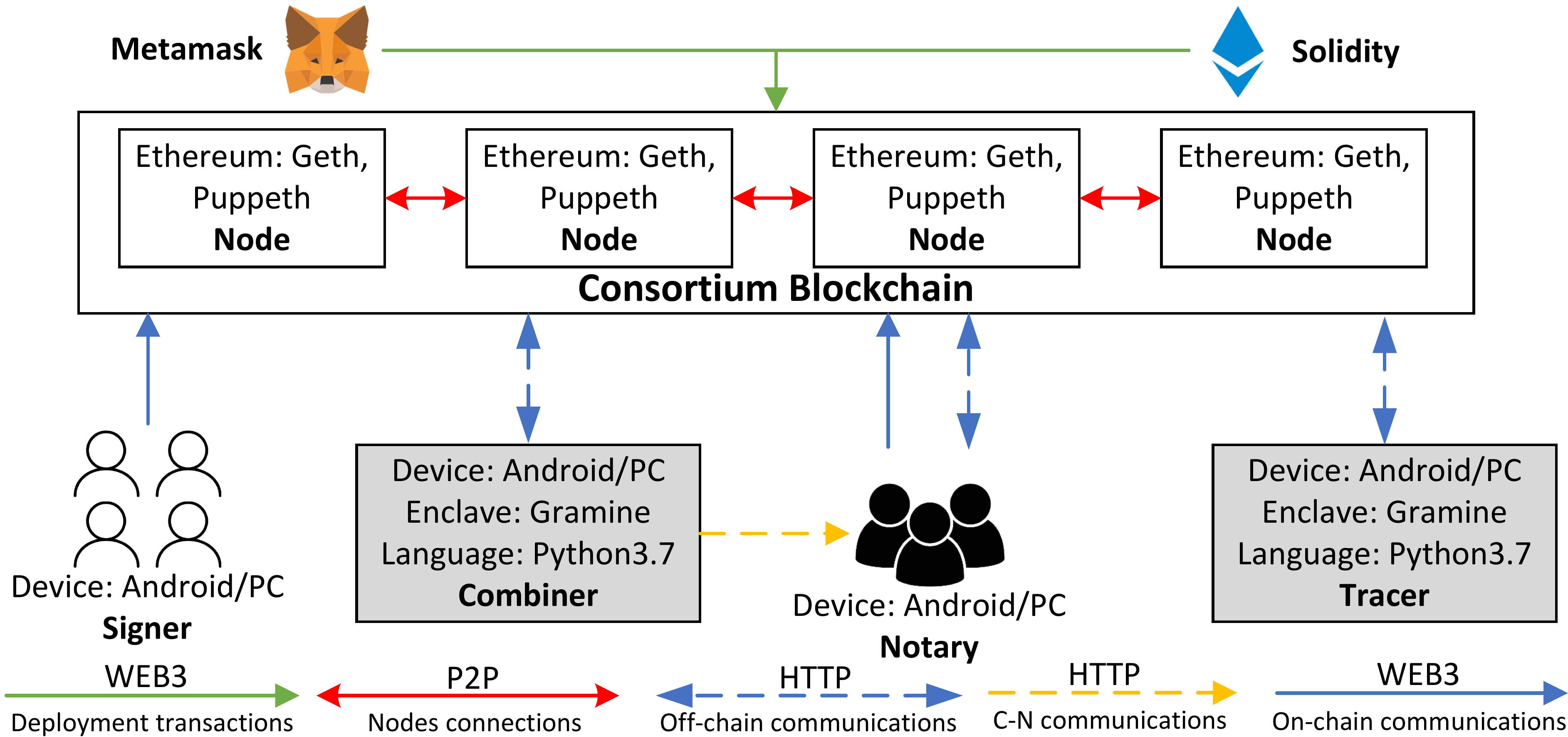}
\caption{Implementation Details of DeTAPS.}
 \label{fig6}
\end{figure}

\vspace{-0.7cm}
\textbf{Setup}.
We implement DeTAPS on a Linux server running Ubuntu 20.04 with a  Intel(R) Xeon(R) Platinum 8369B CPU @ 2.70GHz   processor and 4 GB RAM.
We use \textsf{HMAC-SHA256} as the pseudo-random function to implement the hash functions. We use \textsf{AES} as the symmetric encryption.
We use Geth as the primary tool for Ethereum network environment establishing.
We use remix to  write the SC and deploy it by a light-weighted browser plugin metamask.
We use puppeth to create the genesis block.
We use Python to implement all cryptographic primitives.
The implementation details are shown in Fig.~\ref{fig6}.

\vspace{-0.7cm}

\vspace{-1cm}
\subsection{Computational Cost}
In Setup, DeTAPS generates all keys.
In Signing, a signer computes a signature share.
In Combining, a combiner combines a signature from $t$ signature shares.
In Verifying, a verifier verifies a threshold signature.
In Tracing, a notary computes a trapdoor, the SC searches on indexes, a tracer traces a threshold signature.
We compute the average consumed time of ten experiments for each figure below.
In Fig.~\ref{fig71}, Setup with $n=50$ and $n_3=50$ is about $177$ ms.
In Fig.~\ref{fig72}, Signing a $10$-KByte message is about $52$ ms.
In Fig.~\ref{fig73}, Combining is around 10 s for a $10$-KByte message, $100$ threshold signatures, and $t=5$, i.e., $500$ signature shares.
In Fig.~\ref{fig74}, Verifying is around 10 ms for a $10$-KByte message.
In Fig.~\ref{fig75}, Tracing with varying $t$ is about $4.9$ s for the enclave given $100$ threshold signatures, $t=3$, and $t'=5$.
In Fig.~\ref{fig76}, Tracing with varying $t'$ is about $3.89$ s for the enclave given $100$ threshold signatures, $t'=3$, and $t=5$.

\begin{figure*}[!htb]
\centering
\subfigure[Setup]{
\label{fig71}
\includegraphics[width=0.48\columnwidth]{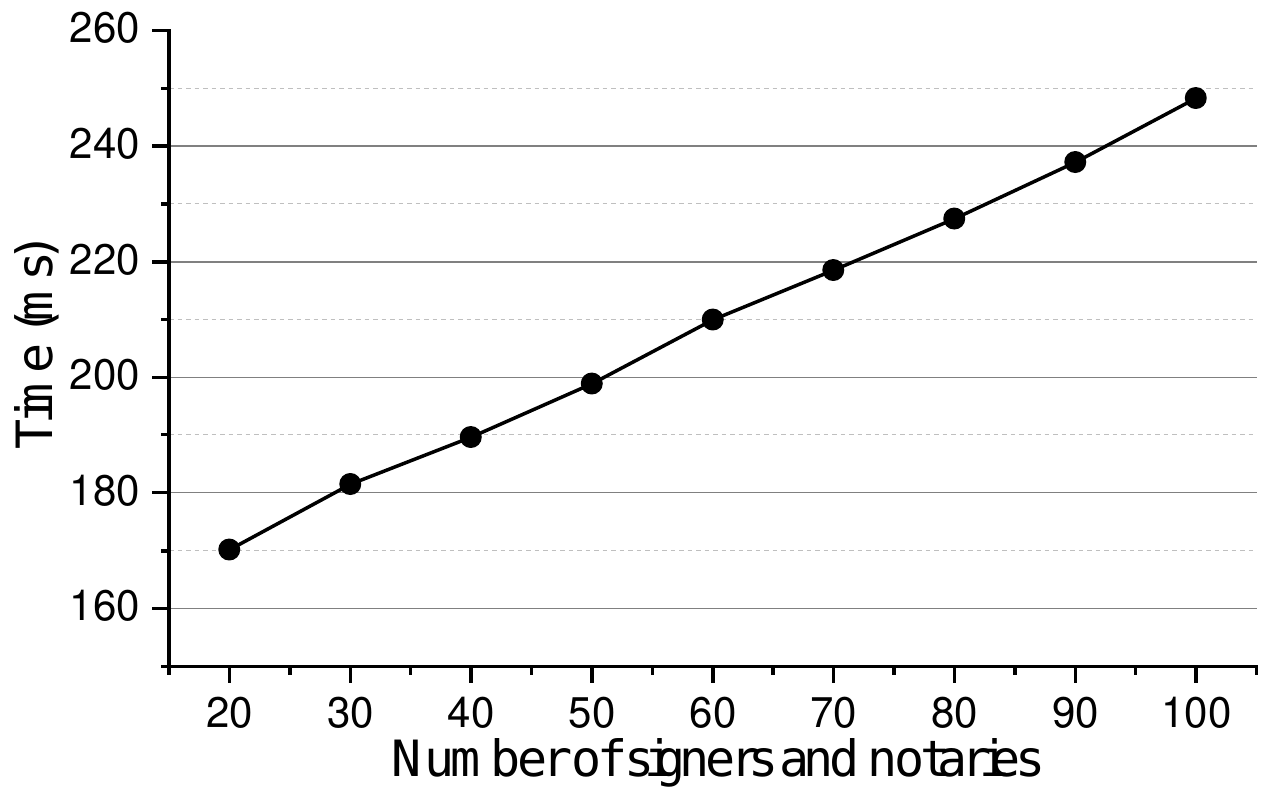}}
\subfigure[Signing]{
\label{fig72}
\includegraphics[width=0.48\columnwidth]{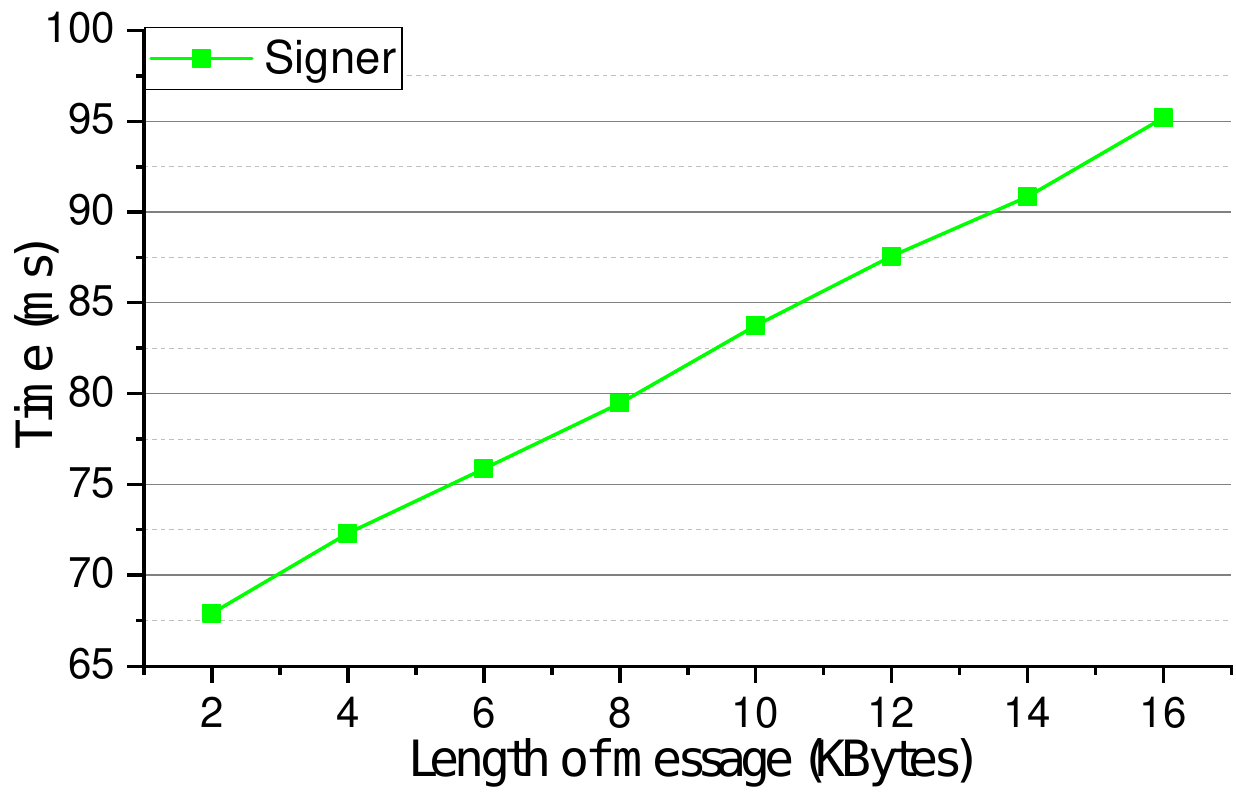}}
\subfigure[Combining]{
\label{fig73}
\includegraphics[width=0.48\columnwidth]{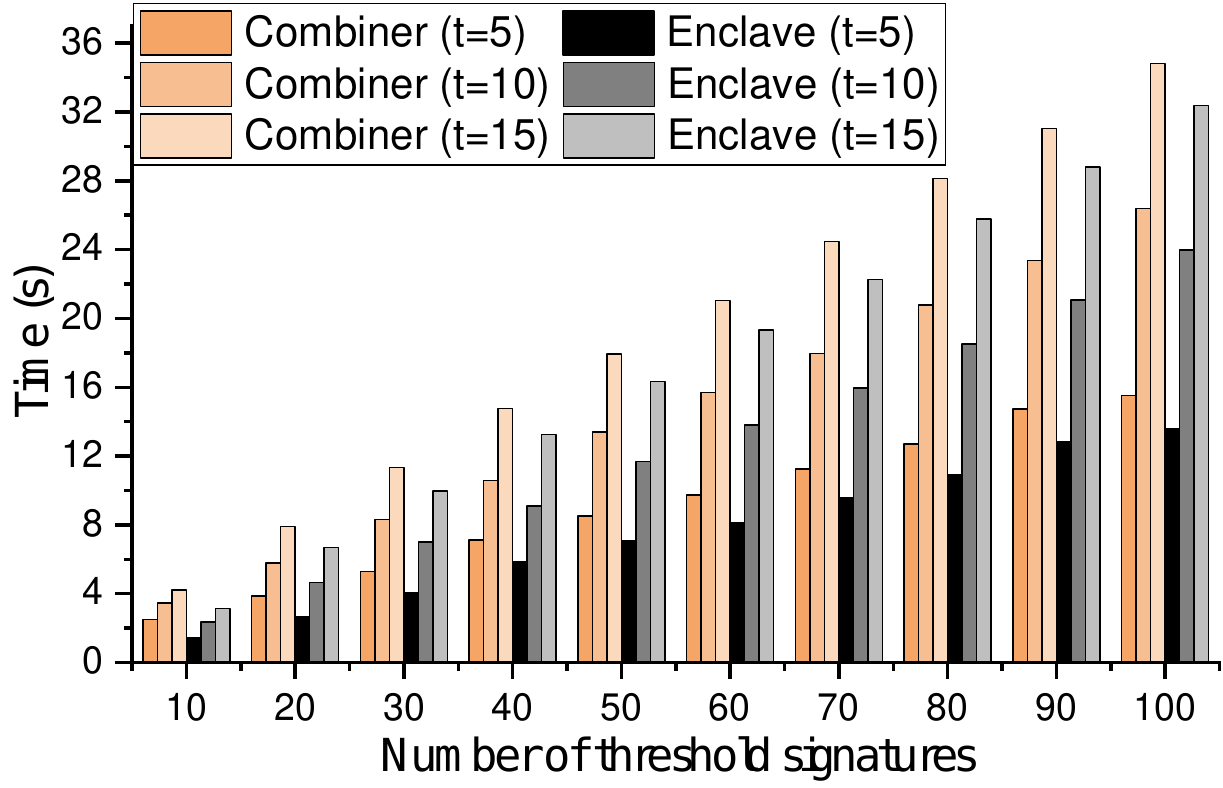}}
\subfigure[Verifying]{
\label{fig74}
\includegraphics[width=0.48\columnwidth]{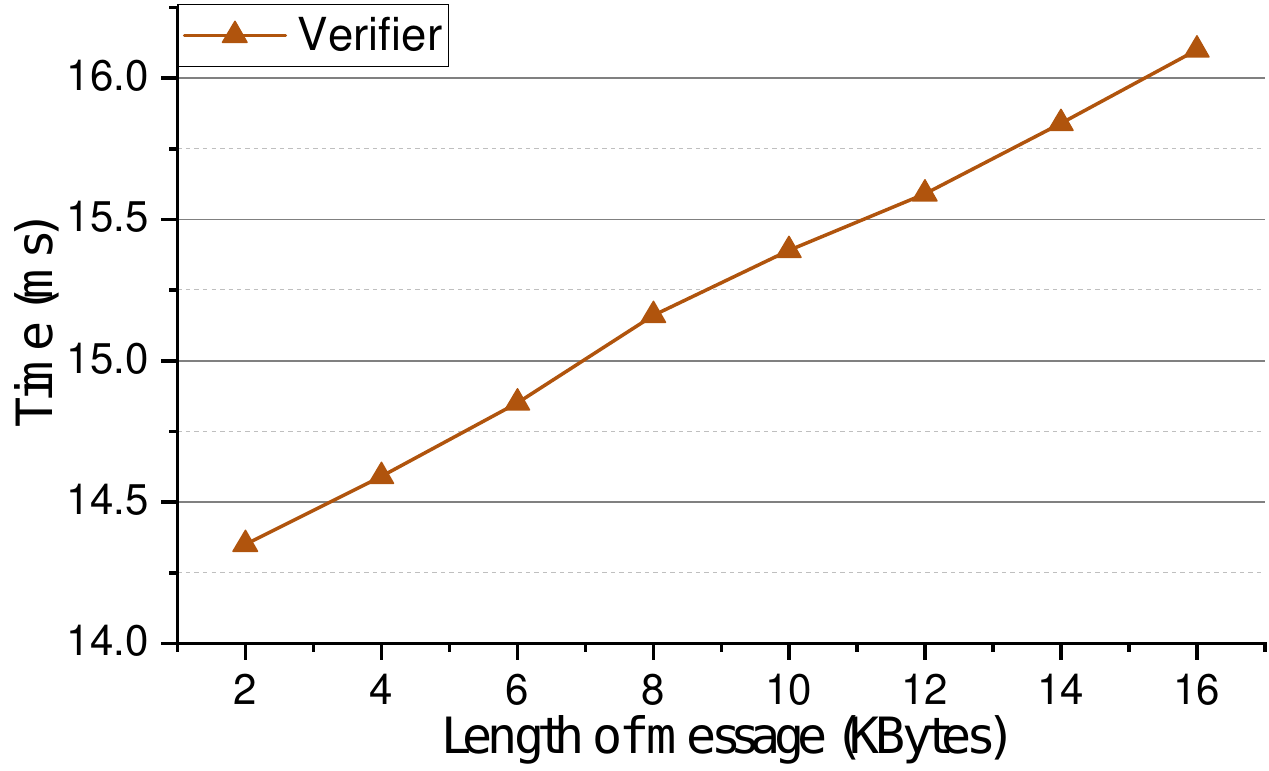}}
\subfigure[Tracing with varying $t$]{
\label{fig75}
\includegraphics[width=0.48\columnwidth]{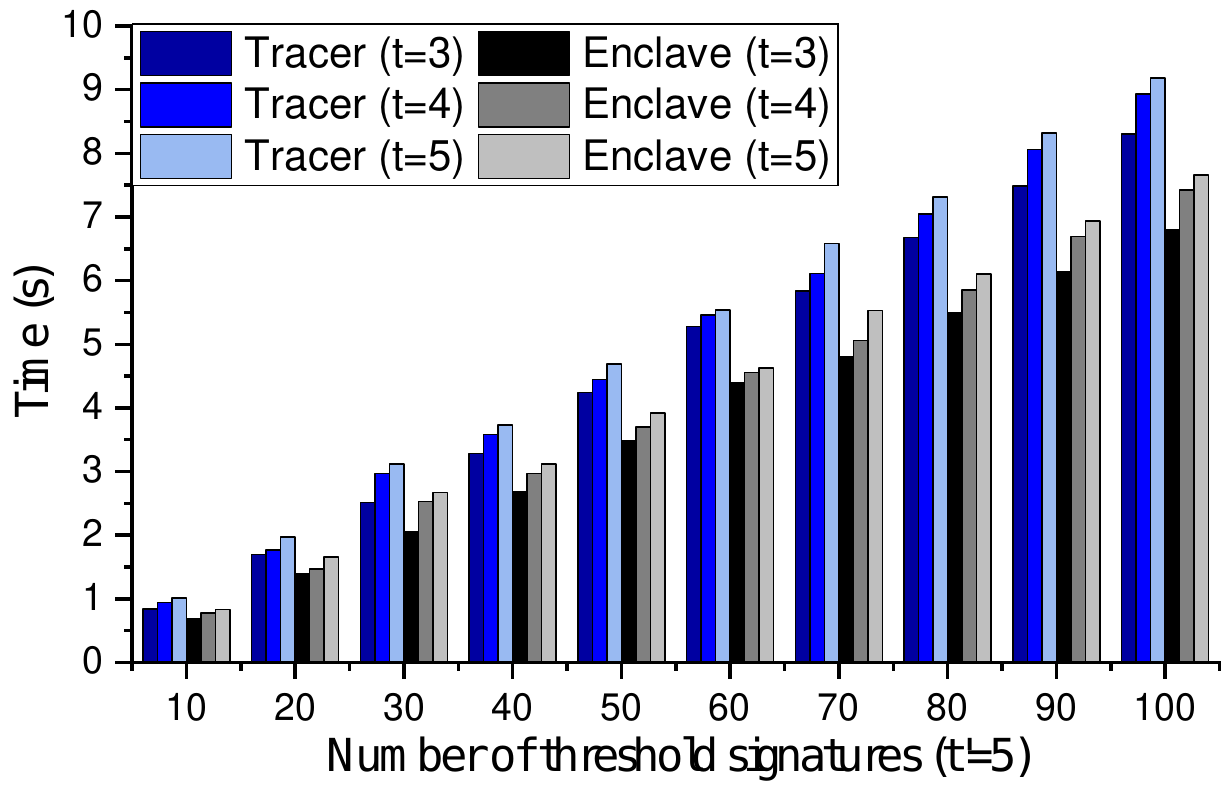}}
\subfigure[Tracing with varying $t'$]{
\label{fig76}
\includegraphics[width=0.48\columnwidth]{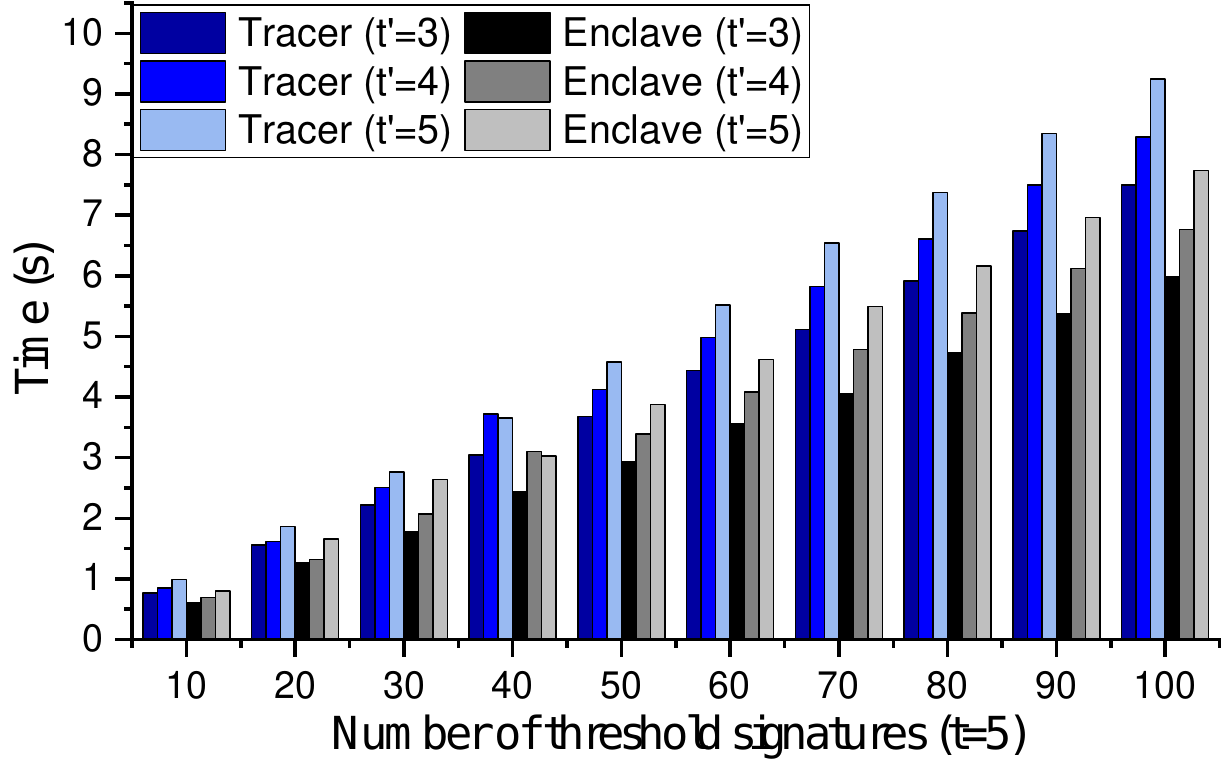}}
\caption{Computational Costs.}
\label{fig7}
\end{figure*}

\subsection{Communication Overhead}
We analyze the communication overhead by counting the length of transmitted messages of all parties for one signing group.
In Signing, a signer sends a signing transaction Tx$^{\textnormal{Sign}}$ including a signature share.
In Combining, an enclave outputs a message $m$, an encrypted threshold signature $\overline{\sigma}$, an encrypted group number $\mathsf{KASE.Enc}(mpk, gid, \mathcal{N})$, and a proof $\pi$. A combiner sends a combining transaction Tx$^{\textnormal{Comb}}$ including $(m, \sigma)$.
In Verifying, the verifier outputs 1 bit.
In Trace, a notary outputs a trapdoor $td$, the SC outputs $t'$ encrypted threshold signatures, the enclave outputs a ciphertext $\mathsf{PKE}.\mathsf{Enc}(\mathcal{S})$, and the tracer relays it to a target party.
We record the communication overhead in Table~\ref{tab3}.

\begin{table*}[h]
\caption{Communication Overhead}
\begin{center}
\begin{tabular}{|m{1cm}<{\centering}|m{1.3cm}<{\centering}|m{1.4cm}<{\centering}|m{1.4cm}<{\centering}|m{1.3cm}<{\centering}|m{1.2cm}<{\centering}|m{1.4cm}<{\centering}|m{1.2cm}<{\centering}|m{1.2cm}<{\centering}|} \thickhline
Phase      &  Signing                                    & \multicolumn{2}{c|}{Combining}                                        & Verifying & \multicolumn{4}{c|}{Tracing}\\ \thickhline
Party       &    Signer                                    & Enclave                               & Combiner                                 &  Verifier     &  Notary & SC & Enclave & Tracer \\ \thickhline
Record  & $1.34$ MB                             & $17.63$ KB                        & $18.06$ KB    & 1 bit              & $0.06$ KB & 1.51$t'$ KB & $2.57$ KB & $2.57$ KB \\ \thickhline
\end{tabular}
\label{tab3}
\end{center}
\end{table*}

\section{Conclusions}
In this work, we have presented DeTAPS, a new threshold signature scheme that achieves unforgeability, accountability, and privacy.
DeTAPS takes a step further towards providing strong privacy as well as notarized and dynamic tracing in a distributed network.
In DeTAPS, the signature threshold $t$ is hidden from distributed combiners and tracers by using an enclave to secure the combining and tracing.
We formally proved the security and privacy of DeTAPS.
Experimental results showed that DeTAPS is efficient, e.g., combining (tracing) a threshold signature for 5 singers (notaries) in the enclave is $86\ (38)$ ms.

\vspace{-0.5cm}
\section*{Acknowledgment}
This work is supported by
National Natural Science Foundation of China (NSFC) under the grant No. 62002094,
National Natural Science Foundation of China (NSFC) under the grant No. 62172040, No. 61872041, No. U1836212, and
National Key Research and Development Program of China under the grant No. 2021YFB2701200, No. 2022YFB2702402.

\begin{itemize}
\item choose randomly $\alpha_i\overset{\$}{\leftarrow} \mathbb{Z}_q, 1\leq i \leq n_3$, compute $B = \prod_{i=1}^{n_3} pk_i^{\alpha_i}$

\item compute $H = \mathsf{hash}(pk_1, \cdots, pk_{n_3}, V, B)$

\item send $(B, \alpha'_1 = b_1H+\alpha_1, \cdots, \alpha'_{n} = b_nH+\alpha_{n})$ to the verifier

\end{itemize}

\textit{Verifier:}

\begin{itemize}
\item compute $H = \mathsf{hash}(pk_1, \cdots, pk_{n_3}, V, B)$

\item check $V^H\cdot B\overset{?}{=} \prod_{i=1}^{n_3} pk_i^{\alpha'_i}$
\end{itemize}

\textbf{1.2 Prove $V_0 = g^{\psi}$ and $V_1 = g^{\sum_{i=1}^{n_3} b_i}\cdot h^{\psi}$:}

\textbf{1.2.1 Prove $V_0 = g^{\psi}$}

\textit{Prover:}

\begin{itemize}

\item choose randomly $\alpha\overset{\$}{\leftarrow} \mathbb{Z}_q$, compute $B = g^\alpha$

\item compute $H = \mathsf{hash}(g, V_0, B)$

\item send $(g, V_0, B, \alpha' = \psi H+\alpha)$ to the verifier

\end{itemize}

\textit{Verifier:}

\begin{itemize}

\item compute $H = \mathsf{hash}(g, V_0, B)$

\item check $V_0^HB\overset{?}{=} g^{\alpha'}$
\end{itemize}

\textbf{1.2.2 Prove $V_1 = g^{\sum_{i=1}^{n_3} b_i}\cdot h^{\psi}$:}

\textit{Prover:}

\begin{itemize}
\item choose randomly $\alpha_i\overset{\$}{\leftarrow} \mathbb{Z}_q, 1\leq i \leq n_3+1$, compute $B = \prod_{i=1}^{n_3} g^{\alpha_i}\cdot h^{\alpha_{n_3+1}}$

\item compute $H = \mathsf{hash}(g, h, V_1, B)$

\item send $(B, \alpha'_1 = b_1H+\alpha_1, \cdots, \alpha'_n = b_{n_3}H+\alpha_{n_3}, \alpha'_{n_3+1}=\psi H+\alpha_{n_3+1})$ to the verifier

\end{itemize}

\textit{Verifier:}

\begin{itemize}
\item compute $H = \mathsf{hash}(g, h, V_1, B)$

\item check $V_1^H\cdot B\overset{?}{=} \prod_{i=1}^{n_3} g^{\alpha'_i}\cdot h^{\alpha'_{n_3+1}}$
\end{itemize}

\textbf{1.3 Prove $b_i(1-b_i) = 0$ for $i=1, 2, \cdots, n_3$: page 67 in Guaranteed Correct Sharing of Integer Factorization with Off-line Share-holders, PKC'98:}

Common input: $Com, g, h\in \mathbb{G}$, Prover's input: $r\in \mathbb{Z}_q$

To prove either $Com = h^r$ or $Com = gh^r$

\textit{Prover:}

if $Com = h^r$

\begin{itemize}
\item choose randomly $w, r_1, c_1\in \mathbb{Z}_q$

\item compute $A = h^w$, $B = h^{r_1}(Com/g)^{-c_1}$, and $H = \mathsf{hash}(Com, A, B)$

\item send $(Com, A, B, c_1, c_2 = H-c_1, r_1, r_2 = w+zc_2)$ to the verifier
\end{itemize}

else if $Com = gh^r$

\begin{itemize}
\item choose randomly $w, r_2, c_2\in \mathbb{Z}_q$

\item compute $A = h^{r_2}Com^{-c_2}$, $B = h^w$, and $H = \mathsf{hash}(Com, A, B)$

\item send $(Com, A, B, c_1=H-c_2, c_2, r_1=w+rc_1, r_2)$ to the verifier
\end{itemize}

\textit{Verifier:}

\begin{itemize}
\item check $H\overset{?}{=} c_1 + c_2\ \textnormal{mod}\ q$

\item check $h^{r_1} \overset{?}{=} B(Com/g)^{c_1}\ \textnormal{mod}\ q$

\item check $h^{r_2} \overset{?}{=} ACom^{c_2}\ \textnormal{mod}\ q$
\end{itemize}

\textbf{2. Prove $\mathsf{ATS.Verify}(pk, m, \sigma^m) = 1$: Sec5.1, Sec5.4, Fig5, Fig6 in TAPS}

\textbf{2.1 Prove $g^z = [\prod_{i=1}^{n} pk_i^{b_i}]^c\cdot R$}

\textit{Prover:}

\begin{itemize}
\item choose randomly $k_z, k_{b1}, k_{b2}, \cdots, k_{bn}\leftarrow \mathbb{Z}_q$, compute $A = g^{k_z}\prod_{i=1}^{n} pk_i^{-c\cdot k_{bi}}$
\item choose randomly $r\leftarrow \mathbb{Z}_q$, compute $B = g^zg^r$, $z' = z+r$, and $R' = Rg^r$
\item compute $H = \mathsf{hash}(pk_1, \cdots, pk_{n}, c, A, B, z', R')$
\item send $(\hat{z} = z'H+k_z, \hat{b}_1=b_1H+k_{b1}, \cdots, \hat{b}_n=b_nH+k_{bn})$ to verifier
\end{itemize}

\textit{Verifier:}

\begin{itemize}
\item compute $H = \mathsf{hash}(pk_1, \cdots, pk_{n}, c, A, B, z', R')$
\item check $A\cdot R'^{H}[\prod_{i=1}^{n}pk_i^{\hat{b}_i}]^c\overset{?}{=} g^{\hat{z}}$
\end{itemize}

\textbf{2.2 Prove $T_0 = g^{\psi}$ and $T_1 = g^{\sum_{i=1}^n b_i}\cdot h^{\psi}$:}

\textbf{2.2.1 Prove $T_0 = g^{\psi}$}

\textit{Prover:}

\begin{itemize}

\item choose randomly $\alpha\overset{\$}{\leftarrow} \mathbb{Z}_q$, compute $B = g^\alpha$

\item compute $H = \mathsf{hash}(g, T_0, B)$

\item send $(g, T_0, B, \alpha' = \psi H+\alpha)$ to the verifier

\end{itemize}

\textit{Verifier:}

\begin{itemize}

\item compute $H = \mathsf{hash}(g, T_0, B)$

\item check $T_0^HB\overset{?}{=} g^{\alpha'}$
\end{itemize}

\textbf{2.2.2 Prove $T_1 = g^{\sum_{i=1}^n b_i}\cdot h^{\psi}$:}

\textit{Prover:}

\begin{itemize}
\item choose randomly $\alpha_i\overset{\$}{\leftarrow} \mathbb{Z}_q, 1\leq i \leq n+1$, compute $B = \prod_{i=1}^{n} g^{\alpha_i}\cdot h^{\alpha_{n+1}}$

\item compute $H = \mathsf{hash}(g, h, T_1, B)$

\item send $(B, \alpha'_1 = b_1H+\alpha_1, \cdots, \alpha'_n = b_nH+\alpha_n, \alpha'_{n+1}=\psi H+\alpha_{n+1})$ to the verifier

\end{itemize}

\textit{Verifier:}

\begin{itemize}
\item compute $H = \mathsf{hash}(g, h, T_1, B)$

\item check $T_1^H\cdot B\overset{?}{=} \prod_{i=1}^{n} g^{\alpha'_i}\cdot h^{\alpha'_{n+1}}$
\end{itemize}

\textbf{2.3 Prove $b_i(1-b_i) = 0$ for $i=1, 2, \cdots, n$: same to 1.3}

\textbf{3. Prove $\mathsf{COM.Verify}(pk, r_{pk}, \textsf{com}_{pk}) = 1$:}

\textit{Prover:}

\begin{itemize}
\item set $A = \textsf{com}_{pk}$

\item choose randmly $\alpha_1, \alpha_2 \in \mathbb{Z}_q$, compute $B = g^{\alpha_1}h^{\alpha_2}$, and $H = \mathsf{hash}(g, h, A, B)$

\item send $(A, B, \alpha'_1 = Hpk+\alpha_1, \alpha'_2 = Hr_{pk}+\alpha_2)$ to the verifier
\end{itemize}

\textit{Verifier:}

\begin{itemize}
\item computes $H = \mathsf{hash}(g, h, A, B)$

\item check $A^HB \overset{?}{=} g^{\alpha'_1}h^{\alpha'_2}$

\end{itemize}

\textbf{4. Prove $\overline{\sigma} = \mathsf{DTPKE.Enc}(ek, \mathcal{N}, \sigma^m)$:}

\ \ \ \ since $c_1 = u^{-k}$ and $c_2 = h^{ks}$, prove similar to 1.2.1
\newline
\newline
\textbf{5. Prove $(c_1, c_2, \{ind_i\})\leftarrow \mathsf{KASE.Enc}(mpk, gid, \mathcal{N})$:}

\ \ \ \ since $ind = e(g, H(pk_i))^t/e(g_1, g_{|\{gid\}|})^t$, $pk_i\in \{\mathcal{N}\}$

\textit{Prover:}

\begin{itemize}
\item set $A =e(g, H(pk_i))/e(g_1, g_{|\{gid\}|})$

\item choose randmly $\alpha\in \mathbb{Z}_q$, compute $B = A^{\alpha}$, and $H = \mathsf{hash}(ind, A, B)$

\item send $(ind, A, B, \alpha' = Ht+\alpha)$ to the verifier
\end{itemize}

\textit{Verifier:}

\begin{itemize}
\item computes $H = \mathsf{hash}(ind, A, B)$

\item check $ind^HB \overset{?}{=} A^{\alpha'}$

\end{itemize}

\end{document}